\documentclass[aps,twocolumn,prc]{revtex4}
\usepackage[dvips]{graphicx}
\usepackage{amsmath}
\usepackage{amssymb}
\newcommand{\be}{\begin{equation}}
\newcommand{\ee}{\end{equation}}
\newcommand{\bea}{\begin{eqnarray}}
\newcommand{\eea}{\end{eqnarray}}
\newcommand{\bsub}{\begin{subequations}}
\newcommand{\esub}{\end{subequations}}

\begin{document}

\title{The outer crust of non-accreting cold neutron stars}


\author{Stefan B. R\"uster}
\email{ruester@th.physik.uni-frankfurt.de}

\author{Matthias Hempel}
\email{hempel@astro.uni-frankfurt.de}

\author{J\"urgen Schaffner-Bielich}
\email{schaffner@astro.uni-frankfurt.de}
\affiliation{
Institut f\"ur Theoretische Physik, J. W. Goethe-Universit\"at, Max von
Laue-Str.~1, D-60438 Frankfurt am Main, Germany}


\date{\today}


\begin{abstract}
The properties of the outer crust of non-accreting cold neutron stars
are studied by using modern nuclear data and theoretical mass tables
updating in particular the classic work of Baym, Pethick and
Sutherland. Experimental data from the atomic mass table from Audi,
Wapstra, and Thibault of 2003 is used and a thorough comparison of
many modern theoretical nuclear models, relativistic and
non-relativistic ones, is performed for the first time. In addition,
the influences of pairing and deformation are investigated.
State-of-the-art theoretical nuclear mass tables are compared in order
to check their differences concerning the neutron dripline, magic
neutron numbers, the equation of state, and the sequence of
neutron-rich nuclei up to the dripline in the outer crust of
non-accreting cold neutron stars.
\end{abstract}


\maketitle


\section{Introduction}
\label{intro}


Neutron stars can be observed as pulsars by their light house effect.
The matter in non-accreting cold neutron stars is in its ground state in
nuclear equilibrium which means that the energy cannot be lowered by
strong, weak, or electromagnetic interactions. Matter in equilibrium
concerning weak interactions is termed $\beta$-equilibrated matter or
matter in $\beta$-equilibrium. Stars are bound by gravity and have to
be charge neutral, otherwise they would be unstable and explode because
of repulsive Coulomb forces.

Neutron stars consist of an atmosphere of electrons, nuclei, and atoms.
Only a fraction of the electrons are bound to nuclei. The ground state
of the nuclei in this regime, with a mass density of
$\rho\lesssim 10^4$~g/cm$^3$, is $^{56}$Fe. The equation of state was calculated by
Feynman, Metropolis, and Teller~\cite{FMT}.

In the context of neutron stars with temperatures above
typically 100~eV, a liquid layer is present between the
atmosphere and the solid crust due to e.g.~hydrogen/helium burning,
where nuclei and electrons are in a liquid phase called the
''ocean''~\cite{ocean}. However, in this work, we focus on
non-accreting neutron stars at zero temperature which is a good
approximation for cold neutron stars.

One assumes complete ionization of the atoms, when the spacing
between nuclei becomes small compared to the Thomas-Fermi radius
$r_\mathrm{TF} \simeq a_0 Z^{- 1/3}$ of an isolated neutral
atom. In this equation, $a_0$ is the Bohr radius and $Z$ the
charge number. The mass density approximately amounts to $\rho
\simeq A m_u n_N$, where $A$ is the mass number, $m_u$ the
atomic mass unit, and $n_N$ the number density of nuclei which
depends on the radius of a spheric nucleus whose volume is the
average volume per nucleus, $4 \pi r_c^3 /3 =
1/n_N$~\cite{Ravenhall}. By combining the last three equations,
one finds that the outer crust of cold neutron stars begins when
$\rho \sim 10^4$~g/cm$^3 \gg 3 A Z$~g/cm$^3$. This shell
consists of nuclei and free electrons. The equation of state was
originally calculated by Baym, Pethick, and Sutherland (BPS)~\cite{BPS}.
The BPS model is valid for zero temperature ($T=0$) which is a good
approximation for the crust of non-accreting cold neutron stars. We will
describe their model of this mass-density regime in more detail in the
next paragraph. The inner crust of neutron stars begins when neutrons
start to drip out of the nuclei at $\rho \simeq 4.3 \times
10^{11}$~g/cm$^3$. This happens because the equilibrium nuclei become more and
more neutron-rich, and finally no more neutrons can be bound to nuclei.
At $\rho \simeq 2 \times 10^{14} $~g/cm$^3$ nuclei do not exist anymore,
signalling the end of the neutron star crust. The equation of state of
the inner crust was calculated by Baym, Bethe, and Pethick~\cite{BBP}
and another equation of state of this regime was derived by Negele and
Vautherin~\cite{NV}. Also a relativistic mean field model has been used
to describe the density regime of the neutron star crust within the
Thomas-Fermi approximation (see~\cite{Shen} and references therein). For
higher densities, the nuclei disintegrate and their constituents, the
protons and neutrons, become superfluid. Muons and hyperons also appear
in this shell. The equation of state in this density regime is
not well known, but it can be modeled by using non-relativistic
many-body theories~\cite{Akmal98} or relativistic nuclear field
theories~\cite{Walecka,RMF,Hanauske00,SZ02}. At extremely high
densities, even the protons, neutrons, and hyperons disintegrate to
their constituents, the quarks. If neutron stars with huge central
densities exist then they can contain a quark core~\cite{Weber} which
probably is color-superconducting~\cite{CSCReviews,QCDphasediagrams}.

In this paper, we focus on the outer crust of non-accreting cold neutron
stars. It contains nuclei and free electrons. The latter become
relativistic above $\rho \sim 10^7$~g/cm$^3$. The nuclei are arranged
in a body-centred cubic (bcc) lattice. The contribution of the lattice
has a small effect on the equation of state but it changes the
equilibrium nucleus to a larger mass number and lowers the total energy
of the system because it will minimize the Coulomb interaction energy of
the nuclei. The latter are stabilised against $\beta$-decay by the
filled electron sea. At $\rho \sim 10^4$~g/cm$^3$, $^{56}$Fe is the true
ground state. With increasing mass density, it is not the true ground
state anymore because the nuclei capture electrons, emit neutrinos and
become neutron richer. When the mass density $\rho \simeq 4.3 \times
10^{11}$~g/cm$^3$, the so-called neutron dripline is reached. Neutrons
begin to drip out of the nuclei and become free. As soon as neutrons
begin to drip out of the nuclei, the outer crust stops and the inner
crust begins.

The composition of the outer crust of non-accreting cold neutron stars
was investigated by Baym, Pethick, and Sutherland (BPS) in a classic
paper 1971~\cite{BPS}. They calculated the equation of state and the sequence
of nuclei which occur in the outer crust of non-accreting cold neutron
stars. They used the nuclear data from the droplet model of Myers and
Swiatecki~\cite{MyersSwiatecki}. The equation of state of the outer
crust of non-accreting cold neutron stars is still commonly taken
nowadays from BPS~\cite{BPS} although it is based on nuclear data of the
mid sixties of the last century. Haensel, Zdunik, and Dobaczewski (HZD)
in 1989~\cite{HZD} used a Skyrme Hartree-Fock-Bogolyubov (HFB) calculation in
spherical approximation for the parameter set SkP~\cite{HFBHZD},
hence ignoring effects from deformations, and the
droplet model from Myers~\cite{Myers} in order to update the results of
BPS. Haensel and Pichon (HP) in 1994~\cite{HP} used the experimental nuclear
data from the atomic mass table of 1992 from Audi and
Wapstra~\cite{AW1992} and the theoretical nuclear mass tables of
the droplet models from M\"oller and Nix~\cite{Moeller}, and Aboussir et
al.~\cite{Aboussir}. A review on the inner and outer crust of accreting as
well as non-accreting neutron stars can be found~\cite{ReviewHaensel}.

In view of the previous work, it seems to be more than timely to
reinvestigate the properties of the outer neutron star crust with
up-to-date and state-of-the-art experimental~\cite{AudiWapstra} and
theoretical mass tables which became available in the last few years via
the Brussels Nuclear Library for Astrophysics Applications
(BRUSLIB)~\cite{BRUSLIB_BCS,BRUSLIB_HFB} and by Dobaczewski and
coworkers~\cite{Dobaczewski} for Skyrme based models and by Geng, Toki and Meng
for a relativistic model~\cite{Geng05}. The nuclear models and their
mass tables used in this work, as listed in detail in
Table~\ref{nuclearmodels}, are taken to update the results of
BPS, HZD, and HP. For the first time, the differences of various
nuclear models concerning the neutron dripline, magic neutron
numbers, the equation of state, and the sequence of nuclei in
the outer crust of non-accreting cold neutron stars are
investigated in detail. To our knowledge, this work is also the
first one which uses mass tables based on modern relativistic
nuclear models. Additionally, effects of pairing and deformation
of nuclei are studied. We find that the inclusion of
deformations for describing nuclei is crucial in determining the
composition of the neutron star outer crust. 

The results of this work rely on the (unknown) properties of
neutron-rich isotopes up to the neutron dripline. The sequence of
neutron-rich nuclei found in this work are in reach to be
measured by the Facility for Antiproton and Ion Research (FAIR)~\cite{FAIR}
at the Gesellschaft f\"ur Schwerionenforschung (GSI), Darmstadt, by 
TRIUMF's Isotope Separator and Accelerator (ISAC-II)~\cite{TRIUMF} and 
by the Rare Isotope Accelerator project (RIA)~\cite{RIA}. 
Detailed experimental determinations of the binding
energy of hitherto unknown nuclei towards the dripline will finally pin
down the actual sequence of neutron-rich isotopes in the outer crust of
neutron stars. Also, the low density equation of state as well as its
composition serves as an important ingredient for low mass neutron star
models as well as for neutron star mergers and core-collapse
supernovae. 

This paper is organised as follows: In Sec.~\ref{model}, we describe how
to find the equilibrium nucleus and how to derive the equation of state
for the outer crust of a non-accreting cold neutron star by using the
BPS model~\cite{BPS}. We use natural units (in units of MeV) and set
$\hbar=c=1$ for the equations presented in the following. We finally
convert our results to the cgs-system in order to be able to compare and
check the results with BPS. In Sec.~\ref{description}, we describe the
nuclear models and their mass tables used in this paper. In
Sec.~\ref{results}, we present our results comparing the various different
modern nuclear models and mass tables used here and compare our findings
to the previous work of BPS, HZD, and HP~\cite{BPS,HZD,HP}. We also show
the differences of the theoretical nuclear models concerning the neutron
dripline, magic neutron numbers, the equation of state, and the sequence
of nuclei in the outer crust of non-accreting cold neutron stars. In
Sec.~\ref{conclusions}, we summarize our results.


\section{The BPS model}
\label{model}


In order to find the equilibrium nucleus and calculate the equation of
state of the outer crust of non-accreting cold neutron stars by using
the BPS model~\cite{BPS}, one has to treat the pressure $P$ as the
independent variable. We start with $P=9.744 \times 10^{18}$~dynes/cm$^2$
which corresponds to the mass density $\rho \simeq 1.044 \times
10^4$~g/cm$^3$. Actually, this special value of the pressure is also the
starting value of the pressure in Ref.~\cite{BPS} and the corresponding
mass density is in good agreement with the approximation made in
Sec.~\ref{intro}: $\rho \sim 10^4$~g/cm$^3 \gg 3 A Z$~g/cm$^3$, the mass
density at which the outer crust begins. Because the pressure in the
star is increasing continuously with decreasing star radius, we increase
$P$ until neutron drip is reached, i.e.~when the chemical potential of baryons
is equal to the neutron mass, $\mu_b=m_n$. With given pressure $P$, we
vary the mass number $A$, and the charge number $Z$ and solve the
equation of the total pressure, \be
\label{pressure}
P = P_e + \frac13 W_L n_N \; ,
\ee
for the electronic density $n_e$. The highest contribution to
Eq.~(\ref{pressure}) is the pressure of free electrons,
\be
P_e = \frac{1}{3 \pi^2} \int_0^{k_e} \frac{k^4}{E_e} dk \; ,
\ee
where $E_e=(k^2 + m_e^2)^{1/2}$.
The electron pressure depends on the electron Fermi momentum
which is related to the electron density, 
\be
n_e = \frac{k_e^3}{3 \pi^2} \; .
\ee
The lattice energy is given by
\be
W_L = - 1.81962 \, \frac{Z^2 e^2}{4 \pi \epsilon_0 a} \; .
\ee
It has the form of the Coulomb energy with a special prefactor
calculated in~\cite{CHM} which arises because of the bcc lattice. 
The bcc lattice constant $a$ is related to the number density of
nuclei by
\be
n_N a^3 = 2 \; .
\ee
The latter one depends on the number density of electrons
because of the neutrality condition which has to be fulfilled in
stars,
\be
n_e = Z n_N \; .
\ee
The baryon density is related to the number density of nuclei,
\be
n_b = A n_N \; .
\ee
The total energy density is given by
\be
\label{Etot}
E_\mathrm{tot} = n_N \left( W_N + W_L \right) + E_e \; .
\ee
The prefactor $1/3$ in Eq.~(\ref{pressure}) originates from the
fact that the pressure of the bcc lattice $P_L$ is one third of the
energy density of the bcc lattice, $P_L/3 = E_L = W_L n_N$. The
energy of the nuclei is obtained by
\be
\label{mass}
W_N = m_n \left( A - Z \right) + m_p Z - b A \; ,
\ee
where $m_n$ is the neutron and $m_p$ the proton mass, and $b$ is
the binding energy per nucleon. The energy density of free
electrons amounts to
\be
E_e = \mu_e n_e - P_e \; ,
\ee
where
\be
\mu_e = \sqrt{k_e^2 + m_e^2}
\ee
is the chemical potential of the electrons and $m_e$ the
electron mass. The quantity to be minimized at fixed pressure
$P$ by varying $A$ and $Z$ is the baryon chemical potential
\be
\mu = \frac{E_\mathrm{tot} + P}{n_b} = \frac{W_N + \frac43 W_L
+ Z \mu_e}{A} \; .
\ee

This procedure has been done for different nuclear models. All of
them contain data in tabular form for $A$, $Z$, or $N = A -
Z$, and the corresponding binding energy $B$ or binding
energy per nucleon $b = B/A$. We use the data of the nuclear
models listed in Table~\ref{nuclearmodels}.

The places at which a phase transition from one to another equilibrium
nucleus happens can be found by varying the pressure as long as its
difference to the pressure of the precise point of the transition
becomes small. As the transition from one nucleus to the next one takes
place, $P$ and $\mu$ of both phases are equal but there will be jumps
in $\mu_e$ and $n_e$ when the proton numbers $Z$ of both phases are
unequal. The baryon density $n_b$ and the mass density 
\be
\rho=\frac{E_\mathrm{tot}}{c^2} 
\ee 
will jump accordingly and are approximately given by 
\bsub
\label{jumps}
\bea
n_b' - n_b &\simeq& n_e \left( \frac{A'}{Z'} - \frac{A}{Z}
\right) \; , \\
\frac{\Delta \rho}{\rho} &\simeq& \frac{\Delta n_b}{n_b} \simeq
\frac{Z/A}{Z'/A'} - 1 \; .
\eea
\esub
The adiabatic index $\Gamma$ is defined by
\be
\Gamma = \frac{n_b}{P} \frac{\partial P}{\partial n_b} \; .
\ee
At the transition point, the adiabatic index jumps to zero
because the pressure in both phases is equal.


\section{Description of the nuclear models used in this paper}
\label{description}


In this section, we describe the nuclear models used in this work as
listed in Table~\ref{nuclearmodels}. For comparison, we add the results
from BPS which were derived by using mass tables from a droplet mass
formula from the sixties.

The most recent mass table of the finite range droplet model
(FRDM)~\cite{Moller97} lists 8979 nuclei ranging from $^{16}$O to $^{339}136$
extending from the proton dripline to the neutron dripline. The mass
table of FRDM is based on a macroscopic finite range droplet model
including a folded Yukawa single particle potential. It gives so far the
best parameterization of masses of known nuclei throughout the nuclear chart.

Non-relativistic Skyrme parameterizations are the well known and widely
used SkM$^\star$~\cite{Bartel82}, SkP~\cite{Doba84} and SLy4
parameter sets~\cite{HaenselSLy4}. Set SkM$^\star$ originates
from the parameter set SkM~\cite{Kri80} which is fitted to
nuclear matter properties and properties of nuclei. Set
SkM$^\star$ is corrected for the systematically too high
binding energies and too low fission barriers of set SkM. The set SkP
derives from Skyrme-Hartree-Fock-Bogolyubov (Skyrme-HFB) calculations
with particular emphasis on pairing effects and the description of
neutron-rich nuclei. The parameters are fitted to nuclear matter
properties, in particular the symmetry energy and properties of
$^{16}$O and $^{208}$Pb to fix the surface energy. Also set SLy4 is
derived to describe in particular neutron-rich isotopes and the
(theoretical) neutron matter equation of state of Wiringa et
al.~\cite{Wiringa88} in order to improve the isospin property away from the
$\beta$-stability line. Nuclear matter properties, the neutron matter
equation of state~\cite{Wiringa88}, and the binding energies and radii
of the doubly magic nuclei $^{40,48}$Ca, $^{132}$Sn, and $^{208}$Pb were
utilized for the fit. Mass tables of these parameter sets were performed
by Dobaczewski et al.~for nuclei up to $Z=108$ from the proton to the
neutron dripline within the Skyrme-HFB approach including effects from
deformation and posted at a publicly available web page
(see~\cite{Dobaczewski}).

The other Skyrme parameterizations are taken from the BRUSLIB
web pages~\cite{BRUSLIB_BCS,BRUSLIB_HFB} and are commonly derived by fitting the
masses of about 2000 known nuclei including effects from deformations.
Different approximations schemes have been used, however. Sets SkSC4 and
SkSC18 use the Extended Thomas-Fermi plus Strutinsky Integral (ETFSI)
approximation for the actual calculation of nuclei. The macroscopic part
is described by an extended Thomas-Fermi approximation, the shell
corrections included by the Strutinsky integral method, and the pairing
energy given by the BCS approximation. The binding energy of about 1700
nuclei were fitted to generate mass tables up to $Z=130,115$,
respectively~\cite{BRUSLIB_BCS}. The set MSk7 originates from a
Skyrme-Hartree-Fock calculation with a ten parameter Skyrme force along
with a four parameter $\delta$-function pairing force using the BCS
approximation for the pairing energy. Its mass table extends to
$Z=120$~\cite{BRUSLIB_BCS}. The sets BSk2, and BSk8, however, are generated by a
full Skyrme-HFB calculation. The binding energies of 2149 nuclei were
fitted for these latter two sets using the experimental mass table of
2003~\cite{AudiWapstra}. The corresponding mass table contains 9200
nuclei up to $Z=120$ lying between the neutron and the proton
driplines~\cite{BRUSLIB_HFB}. The newest parameter set along this line, dubbed
BSk9, has been constrained by fixing the symmetry energy to 30
MeV~\cite{BRUSLIB_HFB} and is not considered in the following. All Skyrme
based mass tables used in this work take into account effects from
deformations. We take the sets SLy4 and BSk8 as state-of-the-art and the
most representatives ones for cross comparison to the other approaches
used in this work (FRDM and relativistic models).

\begin{table}[ht]
\begin{center}
\begin{tabular}{|l l l l|}
\hline
Model & Set & Comments & Refs. \\
\hline
\hline
Droplet
& BPS  & used by BPS~\cite{BPS} & \cite{MyersSwiatecki} \\
& FRDM & Finite Range Droplet Model & \cite{Moller97} \\
\hline
Experiment & & Atomic mass table 2003 & \cite{AudiWapstra} \\
\hline
Non-          & BSk2        & Skyrme HFB & \cite{BRUSLIB_HFB} \\
relativistic  & BSk8        & Skyrme HFB & \cite{BRUSLIB_HFB} \\
& MSk7        & Skyrme HF + BCS & \cite{BRUSLIB_BCS} \\
& SkM$^\star$ & Skyrme HFB & \cite{Bartel82,Dobaczewski} \\
& SkP         & Skyrme HFB & \cite{Doba84,Dobaczewski} \\
& SkSC4       & ETFSI method + BCS & \cite{BRUSLIB_BCS} \\
& SkSC18      & ETFSI method + BCS & \cite{BRUSLIB_BCS} \\
& SLy4        & Skyrme HFB & \cite{HaenselSLy4,Dobaczewski} \\
& SLy4HO      & Skyrme HFB & \cite{HaenselSLy4,Dobaczewski} \\
\hline
Relativistic
& Chiral & Chiral effective model &
\cite{Papa99}-\cite{Schramm03} \\ 
& NL3    & Nuclear field theory & \cite{Lala97,Lala99} \\
& NL-Z2  & Nuclear field theory & \cite{Bender99} \\
& PCF1   & Point coupling model & \cite{Buervenich02} \\
& TMA    & Nuclear field theory & \cite{Geng05} \\
\hline
\end{tabular}
\end{center}
\caption{The nuclear models used in this paper.}
\label{nuclearmodels}
\end{table}

Relativistic nuclear field theories used here are based on the exchange
of mesons or relativistic point-couplings between nucleon fields in the
mean-field and the no-sea approximation. The effective Lagrangians
behind the parameter sets NL3, NL-Z2, and TMA contain the exchange of
scalar, vector, and isovector mesons. Sets NL3 and NL-Z2 include scalar
selfinteraction terms, set TMA in addition vector selfinteraction terms
in the effective Lagrangian. The parameter set NL3 was developed in
particular to describe isospin effects. Besides binding energy and
charge radii of 10 nuclei, neutron radii were included into the fit
procedure~\cite{Lala97}. A selfconsistent microscopic correction to the
spurious center-of-mass motion is performed for the set NL-Z2. Its fit
encompasses binding energies, diffraction radii, surface thicknesses,
charge radii and spin-orbit splittings from a total of 17
nuclei~\cite{Bender99}. The recent parameter set TMA~\cite{Geng05} updates the
sets TM1 and TM2~\cite{Suga94}, which were fitted to binding energies
and charge radii for low mass numbers (TM2) and high mass numbers (TM1).
The fit parameters of the set TMA are chosen to be mass number dependent
so as to have a good description of the properties of light and heavy
nuclei~\cite{Geng05}. Note that the vector field selfinteractions result
in a (vector) selfenergy of the nucleon, which is similar to
Relativistic Brueckner-Hartree-Fock calculations, therefore mimicking
many-body effects of higher order beyond the usual relativistic
mean-field description (see e.g.~\cite{Suga94} and references therein).
The relativistic point coupling model, developed in~\cite{Buervenich02},
consists of four, six, and eight-fermion point coupling terms in the
effective Lagrangian. Its parameter set PCF1 is determined by a list of
similar observables as in the fit for the set NL-Z2 (see above). Pairing
effects are usually included by a standard $\delta$-force within the BCS
approximation. The mass tables for the sets NL-Z2 and PCF1 range from
$Z=26$ to $Z=140$, the one for the set NL3 (in spherical approximation)
from $Z=20$ to $Z=130$ up to several nuclei behind the neutron dripline
to look for particle stable neutron-rich islands of stability. Note that
only even-even nuclei are computed for our purposes as only those nuclei
can possibly appear in the outer crust in nuclear $\beta$-equilibrium.

The chiral effective Lagrangian used for the set denoted as 'Chiral' is
build on the nonlinear realization of the chiral SU(3)$\times$SU(3)
symmetry~\cite{Papa99,Beckmann02,Schramm02,Schramm03} as motivated from
quantum chromodynamics (QCD). The full nonet of scalar, pseudoscalar,
vector mesons is taken into account as well as the baryon octet in the
mean-field and no-sea approximation and a dilaton field. The hadron
masses, meson masses and baryon masses, are not additional input
parameters but are generated by the vacuum expectation values of the
scalar fields of the effective Lagrangian (spontaneous chiral symmetry
breaking besides explicit symmetry breaking) as dictated by the
properties of QCD. Pairing effects are included by the BCS scheme. The
SU(3) chiral model describes nuclear matter as well as properties of
nuclei~\cite{Papa99,Schramm02}, hypernuclei~\cite{Beckmann02} and
neutron stars~\cite{Hanauske00,SZ02,Schramm03}. The mass table used
extends from $Z=16$ to $Z=100$.

For the relativistic models, spherical calculations with and without
pairing effects have been performed for the sets NL3, NL-Z2, PCF1 and
Chiral. Mass tables of calculations including effects from deformations
are publicly available for NL3~\cite{Lala99} and TMA~\cite{Geng05} and
are taken for comparison to FRDM and the Skyrme-based
parameterizations. The mass table for NL3 of~\cite{Lala99} lists 1315
even-even nuclei up to $Z=98$ using the BCS pairing scheme with constant
pairing gaps, while the one for TMA~\cite{Geng05} contains nuclei from
the neutron to the proton dripline up to $Z=100$ with BCS type
$\delta$-force pairing.


\section{Results}
\label{results}


In this section, we present our results obtained within the model of
Baym, Pethick, and Sutherland (BPS)~\cite{BPS} as the equation of state
and the sequences of nuclei of the outer crust of non-accreting cold
neutron stars. Various nuclear models as listed in
Table~\ref{nuclearmodels} are used for this purpose. As shown in
Sec.~\ref{model}, the binding energy $B$ or rather the binding energy per
nucleon $b$ together with the mass number $A$ and the proton number $Z$
are the input parameters for the BPS model, see Eq.~(\ref{mass}).
Different nuclear models, of course, usually have different binding
energies per nucleon for the same nuclei resulting in different
sequences of neutron-rich nuclei in the outer crust of non-accreting
cold neutron stars. Because most of the binding energies of nuclei with
low mass numbers are known precisely from experiments, we prefer using
the nuclear data from the atomic mass table~2003 from Audi, Wapstra, and
Thibault~\cite{AudiWapstra} whenever possible. In this context, we
mention that we do not take any estimated (non-experimental) data of the
atomic mass table~\cite{AudiWapstra} into account. If the corresponding
nuclei are not listed in the atomic mass table~\cite{AudiWapstra}, we
use the data of theoretical nuclear models for calculating the sequences
of nuclei which are present in the crust of non-accreting cold neutron
stars. There is only one exceptional case: The nuclear data of the
droplet model from Myers and Swiatecki~\cite{MyersSwiatecki} which is
used for calculating the original sequence of nuclei obtained by BPS is
not modified by nuclear data from the atomic mass
table~\cite{AudiWapstra} because we want to compare our new
results with the original ones from BPS~\cite{BPS}. By using the
newest and most modern nuclear models, we update the results
of~\cite{BPS,HZD,HP}.

Besides, we study differences between the theoretical nuclear models
directly. The location of the neutron dripline is of great importance
for our investigations, because their position in the nuclide chart is
decisive if a neutron rich nucleus has a chance to be present in the
outer crust of a non-accreting cold neutron star or not. If the nucleus
is behind the neutron dripline, it is unstable and will emit neutrons
even for $\beta$-equilibrium and large electron fractions. Therefore,
the nucleus can not be present in the sequence of nuclei in the outer
crust. Of course, this is not the case if the nucleus is located before
the neutron dripline.

In this section, we subdivide our results in several categories
in order to have a better structure of the results in our paper. In
Sec.~\ref{Equations_of_State}, we show the equations of state by
using various nuclear models. In Sec.~\ref{Effects_of_Pairing},
we discuss the effect of pairing on the neutron dripline and on
the sequence of nuclei. In
Sec.~\ref{Non-relativistic_Parameterizations}, the
non-relativistic parameterizations of BRUSLIB~\cite{BRUSLIB_BCS,
BRUSLIB_HFB} and Dobaczewski~\cite{Dobaczewski} are compared. In
Secs.~\ref{Neutron_Driplines}
and~\ref{Sequences_of_Nuclei}, we compare relativistic and
non-relativistic models for the neutron dripline and
sequences of nuclei to each other. In
Sec.~\ref{Most_Modern_Nuclear_Models}, we show
the results for the state-of-the-art nuclear models and discuss
their proton numbers in the sequences of nuclei. Finally, in
Sec.~\ref{previous_works}, we make a comparison to previous works.

\subsection{Equations of State}
\label{Equations_of_State}
Fig.~\ref{eos} shows the equations of state, i.e.~the pressure as a
function of the mass density. For low mass densities, the nuclei
appearing in the outer crust of non-accreting cold neutron stars are the
same for all nuclear models because they are given by the experimental
data of the atomic mass table~\cite{AudiWapstra}. Hence, the equations
of state in Fig.~\ref{eos} are almost the same. Only the set BPS
exhibits a different sequence of nuclei, because $^{66}$Ni is not found
as an equilibrium nucleus by using the data from the droplet model from
Myers and Swiatecki~\cite{MyersSwiatecki}. But as one can see, this
little deviation in the sequence of nuclei does not have any noticeable
consequences on the equation of state by comparing to all other graphs.
A closer look reveals tiny jumps in the mass density for constant
pressure. This is not surprising because such a behaviour is predicted by
BPS~\cite{BPS} and explained in Sec.~\ref{model}, see Eqs.
(\ref{jumps}). Only small differences can be seen in the high density
range where the graphs begin to separate. In order to really see these
tiny differences, we zoomed into the high mass density region. The jumps
and the separation of the graphs are clearly recognizable in
Fig.~\ref{eoszoom}. At high mass density, the graphs separate from each other
because we use different nuclear models which have different binding
energies per nucleon for the same nuclei. This leads to different
equilibrium nuclei and different equations of state. The equilibrium
nuclei of known nuclei are marked along the graph. The iron nucleus
$^{56}$Fe is the energetically favoured one until the energy density
reaches $10^7$~g/cm$^3$. Then the sequence of nuclei continues with the
nickel isotopes $^{62}$Ni, $^{64}$Ni, and $^{66}$Ni and jumps then to
the heavier nuclei $^{86}$Kr and $^{84}$Se. Beyond $^{84}$Se and a
energy density of about $10^{10}$~g/cm$^3$, we find that the sequence
depends on the nuclear model and the nuclear mass table used.

In Fig.~\ref{adiabaticindex}, we show the adiabatic index $\Gamma$ as a
function of the mass density. As one can see, there are no noticeable
differences between the graphs of the nuclear models used. At high mass density,
$\Gamma$ asymptotically approaches the value of the relativistic limit,
i.e.~$\Gamma=\frac43$. At the transition points from one equilibrium
nucleus to another, the value of the adiabatic index jumps to zero. For
simplicity, these jumps are not shown in Fig.~\ref{adiabaticindex}.

\subsection{Spherical Relativistic Models and Effects of Pairing}
\label{Effects_of_Pairing}
In Fig.~\ref{rmfpairing}, the effect of pairing in nuclei on the neutron
dripline and on the sequence of nuclei in the outer crust of
non-accreting cold neutron stars is depicted. We compare relativistic
nuclear models with pairing to those without pairing effects for
spherical calculations. The form of the neutron dripline shows some
distinct features (but note for the following discussion that the mass
tables for NL-Z2 and PCF1 start only at $Z=26$). In particular, there
are 'plateaus' and 'walls' visible along the neutron dripline indicating
particular strong shell effects for neutrons as well as protons. In
addition, the neutron dripline does not continuously increase with
increasing neutron number, but exhibits 'peninsulas' of particle stable
isotopes, especially pronounced for the set NL-Z2 (with pairing) around
$N=90-100$ and $Z=32$. Hence, a sequence of stable isotopes with increasing
neutron number stops first at the neutron dripline followed by a region of 
unstable isotopes, but then exists again a sequence of stable isotopes for 
higher $N$. These features were only found because the calculations did not 
stop at the neutron dripline but were performed even for several nuclei 
behind the neutron dripline. We also note, that the effect of pairing 
on the neutron dripline is rather small. The inclusion of pairing does not 
shift the neutron dripline but smoothes out the neutron-dripline.

The effect of pairing on the sequence of nuclei is also rather small.
Differences in the equilibrium nuclei can be only seen at large mass
numbers $A$. For low mass numbers, there is no difference because we
use consistently the data of the atomic mass table~\cite{AudiWapstra}
whenever available. However, pairing has an effect on the sequence of
neutron numbers as it splits those of the point coupling model PCF1 and
the chiral model: from 82 to 80 and 82 in the point coupling model PCF1
and from 70 to 68 and 70 in the chiral model. The reason is that the
effect of pairing leads to the occupation of extra energy levels which
are preferred in comparison to the ones without pairing. The smearing of
energy levels also cause the smoothing effect on the neutron dripline.
Note, that the chiral model does not have the usual magic neutron number
82 but the magic neutron number 70. This effect might be related to the
inclusion of tensor terms in the chiral model which are absent in the
other relativistic models used here.

\subsection{Non-relativistic Parameterizations}
\label{Non-relativistic_Parameterizations}
In Figs.~\ref{bruslib} and \ref{dobaczewski}, we compare
non-relativistic Skyrme-type models. There are no big differences
visible for the neutron dripline and the sequence of nuclei between the
parameter sets SkSC4, which uses the ETFSI method, MSk7, a Hartree-Fock
calculation with BCS pairing, and BSk8, the full HFB
calculation, in Fig.~\ref{bruslib}. The neutron driplines of all models
in Fig.~\ref{bruslib} are next to each other and the sequence of nuclei
are almost the same. Only the set BSk8 has a shift to larger proton
numbers $Z$ in the sequence of nuclei, $\Delta Z = 2$. All
parameterizations shown in Fig.~\ref{bruslib} have magic neutron numbers
50 and 82. On the other hand, the nuclear models of Dobaczewski
et al.~show pronounced differences in the neutron dripline as well as in the
sequences of nuclei, see Fig.~\ref{dobaczewski}. The neutron dripline of
the set SkP is shifted to smaller proton numbers compared to the set
SLy4. The neutron dripline of the set SkM$^\star$ is shifted to even
smaller proton numbers in comparison to the neutron dripline of the set
SkP. The sequence of nuclei of the set SLy4 is shifted to larger proton
numbers in comparison to the other two sets, with $\Delta Z_\mathrm{max}
= 4$. The sequence of nuclei for SkP makes an unusual big jump to the
nucleus ${}^{86}$Fe and has also unusual magic proton numbers 30 and 38
while the other two sets only have the magic proton number 28. However,
all models in Fig.~\ref{dobaczewski} have magic neutron numbers 50 and
82. We conclude, that the details of the approximation, ETFSI, BCS, or
HFB, are not important for the location of dripline and the sequence of
neutron-rich nuclei in neutron star matter, as long as the parameters
are fitted to a similar (extended) set of observables. There are,
however, substantial differences if different sets of observables are
used for the fitting procedure. Note, that all mass tables used here
include effects from deformations.

\subsection{Neutron Driplines}
\label{Neutron_Driplines}
We now compare relativistic and non-relativistic parameter sets and
mass tables in detail, delineating in particular the role of
deformations. In Fig.~\ref{drip}, we show the neutron driplines of all
parameter sets as listed in Table~\ref{nuclearmodels} including the
relativistic parameterizations NL3 and TMA within a calculation
including deformations.

By comparing the relativistic models with and without pairing, one again
recognizes that pairing makes the neutron driplines smoother. The
prominent proton number found in the dripline of the point coupling
model PCF1 with pairing is 40. Its neutron
numbers are 68, 80, and 110 with pairing, and 70, 82, and 112 without
pairing. The corresponding proton numbers of the chiral model are 24,
42, and 46 with pairing, and 24, 38, 42, and 46 without pairing.
Its neutron numbers are 60, 68, and 110 with pairing, and 62, 70 
and 112 without pairing. The proton number found for the model
NL-Z2 with and without pairing is 32. Its neutron
numbers are 68, 82, 102, and 112 with pairing, and 70, 82,
102, and 112 without pairing. NL3 with pairing has many distinctive 
steps in the dripline, the proton numbers are 26, 32, 38 and 40. The
neutron numbers also for NL3 without pairing are 62, 70, 82 and 112.
By comparing spherical with deformed
nuclei, one finds that the effect of deformation is that the neutron
dripline rises steeper and nearly linear. The neutron driplines of the
relativistic nuclear models with deformations, NL3 and TMA, extend from
$(Z,N) \simeq (20,45)$ to $(Z,N) \simeq (50,105)$ in a nearly linear and
direct way in stark contrast to the wiggly neutron driplines of the
spherical relativistic calculations. The set TMA has less pronounced
neutron and proton numbers in the neutron dripline in comparison to the
spherical calculations. Noticeable occupied proton numbers of the model
TMA are 22, 26, and 40 which is unusual, and a magic neutron number is 82. 
In some cases, the neutron dripline steps down before continuing upwards 
for the mass table of set TMA (at $Z=30$, $N=74$ and at $Z=38$, $N=88$). 
The model NL3 with deformations has the proton numbers 30 and 44, the 
neutron numbers are 54, 82 and 86.
Compared to the spherical calculations, the deformed calculations of NL3
and TMA show a rather similar straight behaviour, although the precise
endpoint of a certain sequence of isotopes can differ drastically, in
particular for $Z=28$. It is evident, that the inclusion of deformations
is crucial for the overall shape of the neutron dripline but that the
location depends strongly on the parameterization.

We now discuss the (deformed) Skyrme calculations in comparison. All
neutron driplines of the nuclear models taken from BRUSLIB do not show
big differences. Only the neutron dripline of the model SkSC18 decreases
sometimes to much lower proton numbers $Z$ and then quickly rises again.
We attribute this effect to the approximate treatment of shell
corrections to the binding energy which will be particular important
close to the dripline. All neutron driplines extracted from the BRUSLIB
mass tables range from $(Z,N) \simeq (20,48)$ to $(Z,N) \simeq (50,108)$
in a nearly linear fashion. A noticeable proton number in the models of
BRUSLIB is 30 and the neutron number 82. By comparing the nuclear
models from BRUSLIB with the ones from Dobaczewski et al., one can see
that there are no substantial differences between the location of the
neutron driplines. Only the neutron driplines of the models SkM$^\star$
and SkP are shifted to lower proton numbers compared to the other
parameter sets. The neutron driplines of the sets SLy4 and SLy4HO (set
SLy4 calculated in hydrogen oscillator basis) extends from $(Z,N) \simeq
(20,48)$ to $(Z,N) \simeq (50,108)$ while that one of the set
SkM$^\star$ goes from $(Z,N) = (20,56)$ to $(Z,N) = (46,115)$, and the
one from the model SkP from $(Z,N) = (20,46)$ to $(Z,N) = (50,114)$. By
comparing the neutron driplines of the models from Dobaczewski
et al.~to those from BRUSLIB, one recognizes that there are much more marked 
shells effects in the models from Dobaczewski et al. Noticeable proton
numbers of the models SLy4 and SLy4HO are 32, 42, and 46 and the neutron
number 82. A marked proton shell of the set SkM$^\star$ is 30. 
A marked neutron shell of this set is 82. The
set SkP exhibits the unusually proton number 22 in its dripline, like
the relativistic set TMA. The set SkM$^\star$ exhibits a noticeable downward
sequence of the neutron dripline around $Z=40$ and $N=100$. Similar
features of downward shifts in the neutron dripline are seen in the
sequences of the parameter sets from BRUSLIB around $N=104$ and $N=108$. We
checked that these features are also present for the HFB set BSk8 and
not due to the approximation scheme used (BCS or ETFSI).

The neutron dripline of the finite range droplet model FRDM ranges from
$(Z,N) = (20,50)$ to $(Z,N) = (50,112)$. It has four pronounced proton
numbers: 28, 34, 44 and 48. A noticeable neutron shell is again 82. By
comparing all neutron driplines from Fig.~\ref{drip}, one recognizes
that all non-relativistic models show the magic neutron number 82 to be
present at the neutron dripline and that all proton shell closures of
the non-relativistic models are less noticeable in comparison to it.
These findings are different to the ones for the relativistic models
(even when including deformations) in Fig.~\ref{drip} where both,
proton and neutron shells, are more noticeable.

\subsection{Sequences of Nuclei}
\label{Sequences_of_Nuclei}
In Fig.~\ref{zvonn}, we show the sequences of nuclei in the outer crust
of non-accreting cold neutron stars by using various nuclear mass
tables. By comparing the sequences of nuclei with pairing to those
without pairing, one recognizes that pairing has the effect of splitting
the magic neutron numbers. The magic neutron number 82 is split into
80 and 82 in the chiral model and in the point coupling model PCF1. By
comparing spherical with deformed calculations, one can see that the
effects of deformation on the sequence of nuclei is small. Models with
and without deformation have the same magic neutron numbers: 50 and 82.
They also have the same sequence of neutron-rich nuclei in the nuclear
chart. An exceptional case is the sequence of nuclei of the chiral model
which shows differences concerning the magic neutron numbers. The
sequence of nuclei within the chiral model does not follow the magic
neutron number 82 but instead continues down the neutron number 70
basically all the way to the neutron dripline. Again, we attribute this
behaviour to the inclusion of tensor terms for the $\rho$ meson in the
effective Lagrangian which induce a different isospin dependent
spin-orbit strength compared to the other relativistic models.

The different sequences of nuclei for the parameter sets of BRUSLIB are
nearly on top of each other and are strikingly along the magic neutron
numbers 50 and 82. There a only slight differences, as the sequence for
set BSk8 reaches $Z=46$ at $N=82$, while all other sets from BRUSLIB
start the sequence along $N=82$ at $Z=44$. The set SLy4 features similar
strong correlations in the sequence of nuclei along $N=50,82$, but like
the relativistic sets NL3 and TMA populates at maximum $Z=42$ for
$N=82$. 

By comparing the sets from BRUSLIB and SLy4 to those of
SkM$^\star$ and SkP one recognizes that the sequences of nuclei
from SkM$^\star$ and SkP are distinctly different in their paths
in the nuclear chart as they mainly follow along isotopes and
not along the magic neutron numbers 50 and 82. This finding is in stark 
contrast to the other parameterizations and mass tables used here, making 
the sets SkP and SkM$^\star$ noticeable exceptions from our general 
results. 

In particular, the more modern Skyrme HFB mass tables of 
the sets SLy4 and BSk8 do not exhibit a pronounced sequence along 
isotopes but are morestuck to the magic neutron numbers 50 and 82. 
The general trend of the sequence of nuclei being along these magic neutron 
numbers is supported
by the calculation for the FRDM which is strikingly similar to the
results for the sets TMA and SLy4. Surprisingly, the classic sequence of
BPS, although using rather old mass tables, arrived at a similar
sequence for large mass numbers compared to the most modern mass tables!

\subsection{Neutron Driplines and Sequences of Nuclei of
State-of-the-Art Nuclear Models}
\label{Most_Modern_Nuclear_Models}
In Fig.~\ref{topmodels}, we compare the most modern and state-of-the-art
mass tables of all the sets listed in Table~\ref{nuclearmodels}. There
are no drastic differences in the shape of the neutron dripline. The
neutron driplines of all the sets presented in the figure have an
approximately linear behaviour extending from $(Z,N) \simeq (20,46)$ to
$(Z,N) = (36,82)$. At the magic neutron number 82, the
neutron driplines suddenly change their slope and follow a vertical
path from $(Z,N) = (36,82)$ to $(Z,N) = (40,82)$. Note that at
the magic neutron number 82, the neutron driplines of all the
selected sets are equal. Therefore, the region at $N=82$ and
around $Z=38$ shows to be the one with the smallest difference
in the precise location of the neutron dripline for all modern
sets used here. From $(Z,N) = (40,82)$ to $(Z,N) \simeq
(50,108)$, the neutron driplines again have approximately the same
gradient as for low mass numbers. But there are differences in the regions of the
nuclide chart where $Z \simeq 28$ and $N \simeq 62$: If one compares the
neutron driplines from the mass tables of sets TMA and NL3 (which
includes effects from deformations), then one can see that their neutron
driplines differ by $\Delta Z = 4$ and $\Delta N = 16$. There are also
differences of similar range for the neutron driplines at $(Z,N) \simeq
(44,100)$ if one compares the model TMA with SLy4 and NL3. Also, the
sequences of nuclei in $\beta$-equilibrium of the modern sets shown do
not exhibit pronounced differences, on the contrary, the sequences are
nearly on top of each other. There are only small exceptions, as there
is a shift to larger proton numbers in the sequence of nuclei for the
set BSk8 with $\Delta Z_\mathrm{max} = 4$. The sets BSk8 and NL3 have
one nucleus in their sequence with $N=52$, set NL3 ends its sequence at
$N=84$. Besides that and the starting iron and nickel isotopes, the
sequences of nuclei for all the selected models and mass tables follow
tenaciously the magic neutron numbers 50 and 82 throughout the (unknown)
nuclear chart until hitting the neutron dripline. 

The endpoint of the sequence of nuclei for all the modern mass tables
studied here is quite similar and happens to be at $Z=34-38$ with $N=82$
($N=84$ for the set NL3). In this region of the nuclear chart, the
neutron driplines of the various sets examined here do also demonstrate
to be rather similar. These features most likely originate from the fact
that the magic neutron number 82 prevails up to the neutron dripline in
the nuclear mass tables used here, which however, relies on a
substantial extrapolation from the masses of known nuclei up to the
neutron dripline. Shell quenching effects, as advocated by
HZD~\cite{HZD}, can possibly exist, in particular for different
isospin dependences of the spin-orbit terms. Then the sequence of nuclei as well
as its endpoint can be located quite differently as seen e.g.\ for sets
SkM$^\star$, SkP and Chiral.

Finally, we note that in all our calculations, there do not appear any
super-heavy nuclei in the sequences of nuclei of the nuclear models and
sets which we use in this work. We checked that superheavy elements can
indeed appear in the sequence of neutron-rich nuclei in the outer crust
of neutron stars if we artificially increase their binding energies per
nucleon $b$ by about one MeV.

In Fig.~\ref{topmodelsZvonrho}, we show the dependence of the proton
number $Z$ on the mass density $\rho$ by using the representative models
of Table~\ref{nuclearmodels}. In this plot, we restrict ourselves to the
high mass density region $\rho \ge 10^{10}$~g/cm$^3$ because for
$10^4$~g/cm$^3 \lesssim \rho \lesssim 3.5 \times 10^{10}$~g/cm$^3$ we obtain the
same elements for each set as they are fixed by the experimental nuclear
data of~\cite{AudiWapstra}: $_{26}$Fe, $_{28}$Ni, $_{36}$Kr, $_{34}$Se,
$_{32}$Ge, and $_{30}$Zn. For larger mass densities, the proton numbers
of most of the sets shown are the same: $_{28}$Ni, $_{42}$Mo, $_{40}$Zr,
$_{38}$Sr, and $_{36}$Kr. The only differences for these mass tables is
that they arrive at the corresponding proton numbers at different mass
densities. For the set TMA, $_{34}$Se and not $_{36}$Kr appears at the
highest mass densities in the outer crust of non-accreting cold neutron
stars. The set BSk8 is the only one that shows significant differences
in comparison to the other sets plotted: For mass densities of $\rho
\simeq 10^{11}$~g/cm$^3$, the BSk8 model has quite different proton
numbers as $_{46}$Pd and $_{44}$Ru appear in the sequence of nuclei.

\subsection{Comparison to previous works}
\label{previous_works}
In order to compare the sequences of nuclei of BPS~\cite{BPS},
HZD~\cite{HZD}, and HP~\cite{HP} with our new results as presented
here, we list all the sequences of nuclei of these previous
works in tabular form. Table~\ref{BPStable} shows the sequence
of nuclei of BPS, Tables~\ref{HZDtable1} and~\ref{HZDtable2} the
sequences of nuclei of HZD~\cite{HZD},
Tables~\ref{HaenselPichontable} and~\ref{HaenselPichontable2}
the sequences of nuclei of HP~\cite{HP}. Finally,
Tables~\ref{BSk8table} and~\ref{TMAtable} summarize the
sequences of nuclei obtained in this work by using the modern
nuclear mass tables from BSk8 and TMA as the characteristic ones
of our whole set of mass tables investigated. Note that BPS and
HZD only use theoretically computed mass tables, while we, as
HP, incorporate experimental data tables in addition. As we know
from Fig.~\ref{eos}, the equation of state is not affected
significantly by a different sequence of nuclei. That is why we
do not show the number density, the pressure and the mass
density in every table. Besides that, we are mainly interested
in the sequence of nuclei here. As one can see from
the tables, the nucleus ${}^{66}$Ni is not present in the crust of
non-accreting cold neutron stars if one uses the droplet model from
Myers and Swiatecki~\cite{MyersSwiatecki} or the spherical model
SkM$^\star$ from Dobaczewski, Flocard, and Treiner~\cite{HFBHZD}. By
taking newer experimental data, this nucleus always exists in the crust.
Even with the droplet model of Myers~\cite{Myers}, this nucleus is
found. A second difference in the sequence of nuclei of BPS~\cite{BPS}
compared to HZD, HP, BSk8, and TMA is that in the latter three one,
${}^{76}$Fe never occurs. It only occurs by using the droplet model from
Myers and Swiatecki~\cite{MyersSwiatecki} or that one from
Myers~\cite{Myers}. When we compare the sequence of nuclei of the model BSk8
with the sequence of nuclei from HZD, HP and that from BPS, we recognize
that the nucleus ${}^{78}$Ni is not present in the sequence of nuclei of
the model BSk8. The reason why ${}^{78}$Ni appears in the sequence of
nuclei from HZD and not in the sequence of nuclei of the BSk8 model is
that HZD used the atomic mass table from 1993~\cite{AW1992} and we use
the newer one from 2003~\cite{AudiWapstra}. With the set TMA, one
obtains a similar sequence of nuclei as for the droplet model of
M\"oller and Nix~\cite{Moeller}, see
Tables~\ref{HaenselPichontable} and~\ref{TMAtable}. The only
differences are that by using the set TMA, ${}^{126}$Ru is not
present in the sequence of nuclei and that the sequence ends
with the nucleus ${}^{116}$Se and not with ${}^{118}$Kr.

The sequence of nuclei obtained by HZD~\cite{HZD} using the set SkP within 
a spherical calculation of Dobaczewski, Flocard, and Treiner~\cite{HFBHZD} 
(see table~\ref{HZDtable1}) is drastically different from all other sequences of
nuclei listed here. Unusual nuclei for the crust, such as ${}^{68}$Ni or
later in the sequence, nickel nuclei with large neutron numbers appear. 
In addition, our results for the same set SkP differ
from the results of HZD~\cite{HZD}. Note,
that the mass table used by HZD is based on a spherical HFB calculation
(and a different pairing force) while the one used in this work includes
effects from deformations and a modern $\delta$-force pairing. The
sequences of nuclei for both mass tables follow closely a certain proton
number for a wide range of isotopes, but HZD find $Z=28$ while we find
$Z=30$ to be prominent. Only the set SkM$^\star$ shows a similarly wide
sequence of nuclei with constant $Z$, however appearing for $Z=28$ and
not for $Z=30$. The sequences of nuclei found here differ from the one
by HZD also at large mass numbers. HZD find, that the sequence jumps
from $Z=28$ to $Z=40$ ending with the nucleus ${}^{134}$Zr while the
sequence in our calculation populates $Z=38$ (and once $Z=26$) and ends
with the nucleus ${}^{126}$Sr. We stress, that we compare here different
mass tables which are, although based on the same parameter set SkP,
computed in significantly different approximations (spherical versus
deformed) and use different pairing forces. The general trend of the
sequence of neutron-rich nuclei, however, is rather similar, as the
sequence follows mainly isotope sequences (constant $Z$) and not the
magic neutron numbers 50 and 82. 

The last nucleus in the
sequences of nuclei found by using different theoretical nuclear models
are shown in Table~\ref{last}. All of the models listed in
Table~\ref{last} except NL3 and that of~\cite{HFBHZD} have the magic neutron
number 82. The charge of the final nucleus of the sequence varies
between $Z=32-40$ for previous works while we find a range of $Z=34-38$
for the modern mass tables.


\section{Summary and Conclusions}
\label{conclusions}


In this work, we investigated the outer crust of non-accreting cold
neutron stars by using the BPS model and the newest nuclear data. If
data from the atomic mass table of 2003 from Audi, Wapstra, and
Thibault~\cite{AudiWapstra} was available for the corresponding
nuclei, we always preferred using this experimental data. We
updated previous work of BPS~\cite{BPS}, HZD~\cite{HZD}, and
HP~\cite{HP} which were based on older data and/or mass
parameterizations. We also compared various nuclear models as
listed in Table~\ref{nuclearmodels} in order to check their
differences concerning the neutron dripline, magic neutron numbers, the
equation of state, and the sequence of nuclei in the outer crust of
non-accreting cold neutron stars. To our knowledge, this is the first
detailed investigation of this kind for an enlarged set of most modern
nuclear models and state-of-the-art theoretical mass tables, which
entails pairing effects, includes effects of deformation and studies
relativistic models in comparison to non-relativistic ones.

We obtain the following results: The equation of state is not affected
by small differences in the sequence of nuclei. There are jumps in the
mass density at constant pressure if the equilibrium nucleus of the
ground state changes to another one. The adiabatic index $\Gamma$ is
not affected by changes in the sequence of nuclei and by using data from
different nuclear models. The location of the neutron driplines of the
Skyrme type models SkP and SkM$^\star$ are at smaller proton numbers $Z$
compared to the other Skyrme type models used here (SLy4 and the ones of
BRUSLIB). The neutron dripline rises steeper and is nearly linear in
mass tables including effects from deformation as compared to spherical
calculations. Pairing has the effect of smoothing out steps on the
neutron dripline found when pairing effects are switched off. Besides
that, pairing has an impact on the magic neutron numbers towards the
neutron driplines. It splits the magic neutron number of the point
coupling model PCF1 from 82 to 80 and 82 and the magic neutron number of
the chiral model from 70 to 68 and 70. The sequences of nuclei obtained
by using theoretical nuclear models can have some differences at large
mass numbers $A$ because they predict different binding energies. For
low mass numbers, as we use the experimental data from the atomic mass
table, the sequences of nuclei in the outer crust of neutron stars are
found to be identical. The nucleus ${}^{66}$Ni does not occur in the
sequences of nuclei of BPS and HZD but it is present in the sequences of
nuclei if one uses newer nuclear data. The sequence of nuclei usually
found in all parameterizations used proceeds along the magic neutron
numbers 50 and 82 all the way to the neutron dripline. Only the sets
for SkP and SkM$^\star$ show sequences of nuclei which follow a certain
(magic) proton number, i.e.~$Z=28,30$ and $Z=38$, which is in contrast
to all other sets investigated here, being non-relativistic or
relativistic ones.

By comparing a selected set of modern models and mass tables, the
location of the neutron dripline as well as the sequence of nuclei in
$\beta$-equilibrium are found to be strikingly similar. The sequence of
nuclei follows nearly entirely the magic neutron numbers 50 and 82. The
set BSk8 differs slightly from all the other modern models as the
sequence of nuclei is shifted to larger proton numbers while following
the standard path along $N=50$ and 82. The endpoint of the sequence
coincides with a region in the nuclear chart where the location of the
neutron dripline of the various models investigated here coincides.
Hence, the final nucleus in the sequence can be pinned down in a rather
narrow range and is extracted to be around $Z=36$ and $N=82$. 

Finally, we note that in all our calculations, there do not appear any
super-heavy nuclei in the sequences of nuclei of the nuclear models
which we use in this paper. Some of them occur if the binding energies
per nucleon $b$ is increased artificially by about one MeV.

The results presented here rely only on the binding energy or mass of
neutron-rich isotopes in the range of $Z=26$ to $Z=40$, a region of the
nuclear chart which will be explored in the near future by the FAIR
facility at GSI, Darmstadt, as well as by the ISAC facility at TRIUMF
and by the RIA project. With these measurements, the non-relativistic
and relativistic models utilized in this work will be tightly
constrained on their isospin properties towards the neutron dripline and
one can shed more light on the actual sequence of neutron-rich nuclei in
the outer crust of neutron stars.
We point out that the proton number of nuclei is important for
heat transport and for Ohmic dissipation of the magnetic field.
Other extraterrestrial sources of information might come from
the observations of x-ray binary systems, where a neutron star
is accreting material from its stellar companion. X-ray bursts
are generated by accretion in the outer crust of neutron
stars which transport material of the crust to the surface. The
x-ray satellites XMM-Newton, Chandra or in the near future XEUS
can give a detailed spectroscopic analysis of an x-ray burst.
Cottam, Paerels, and Mendez~\cite{Cottam2002} reported about
redshifted spectral lines of highly ionized elements in x-ray
bursts of the low-mass x-ray binary EXO 0748-676.


\begin{acknowledgments}


We thank Thomas B\"urvenich, Thomas Cornelius, Stefan Schramm,
and Lisheng Geng for providing the mass tables of the relativistic
nuclear models used in this research project. We also thank J.
A. Maruhn and P. G. Reinhard for useful discussions.
S. R. thanks for using the Center for Scientific Computing (CSC) of the
Johann Wolfgang Goethe-Universit\"at. We thank the referee for
useful comments. This work is supported by the Gesellschaft
f\"ur Schwerionenforschung (GSI) in Darmstadt via the program
"Fremde F\&E-Mittel" in the research project OF/SAF "Struktur
von neutronenreichen, exotischen Kernen in der Kruste von
Neutronensternen".

\end{acknowledgments}



\begin{widetext}


\begin{figure}[ht]
    \includegraphics[scale=0.65]{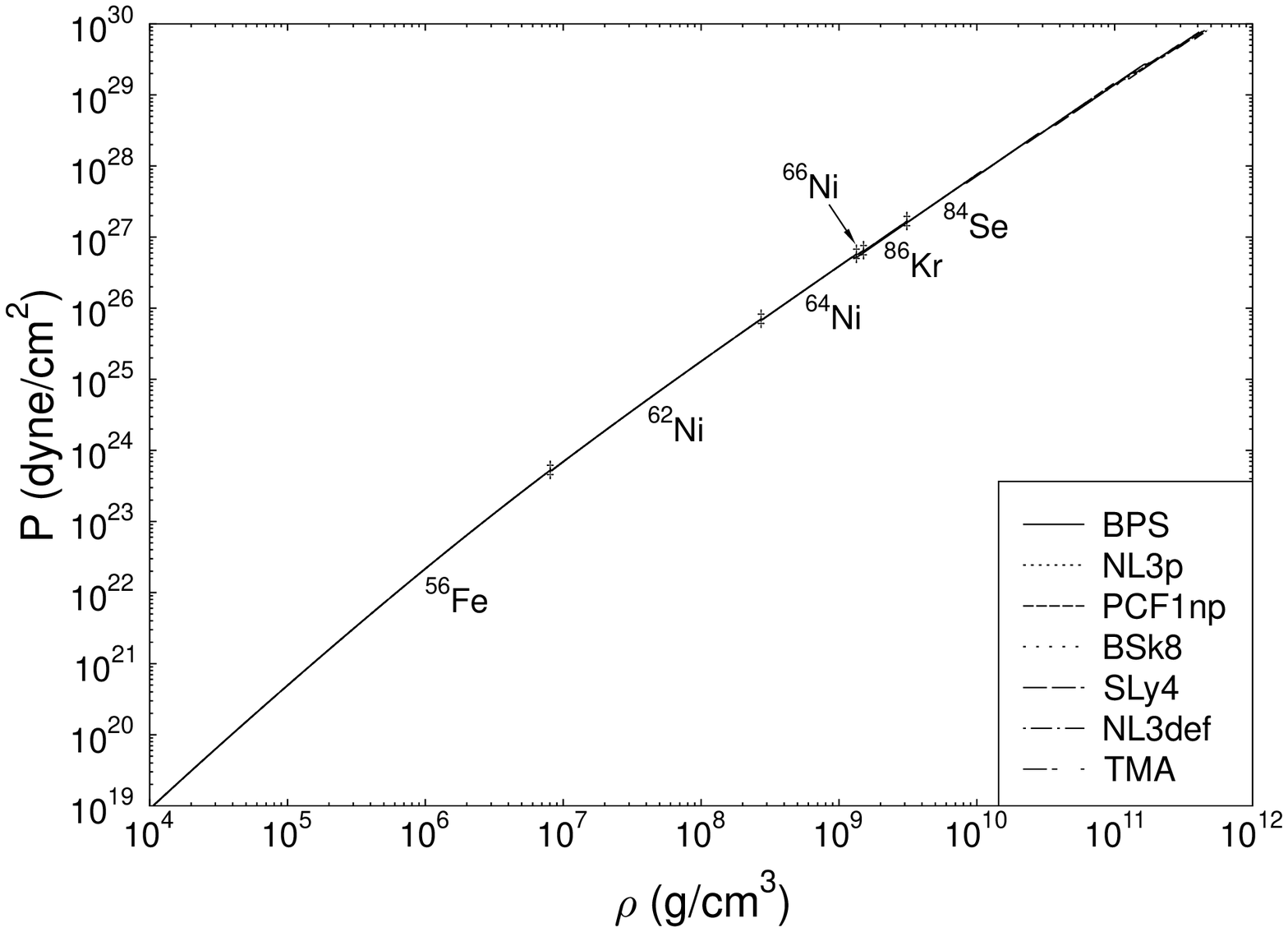}
    \caption{The equations of state, i.e.~the pressures as a
    function of the mass densities, calculated by using the BPS
    model and the binding energies of various nuclear models (NL3p: set
    NL3, spherical calculation with pairing, PCF1np: set PCF1 without
    pairing, NL3def: set NL3 in the deformed calculation).}
    \label{eos}
\end{figure}

\begin{figure}[ht]
    \includegraphics[scale=0.65]{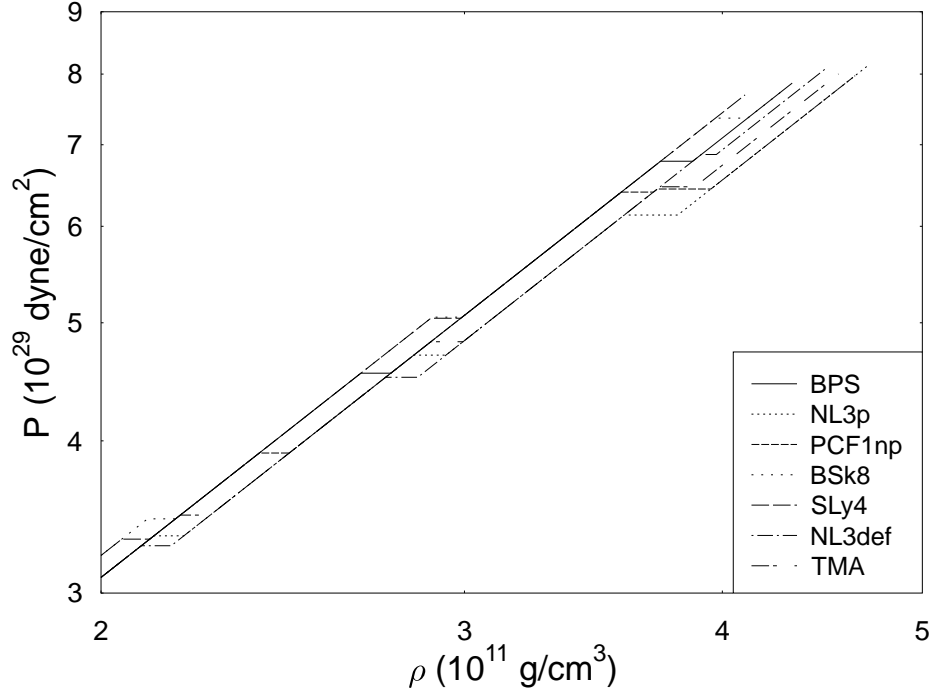}
    \caption{The same as Fig.~\ref{eos} for the high
    mass density range.}
    \label{eoszoom}
\end{figure}

\begin{figure}[ht]
    \includegraphics[scale=0.65]{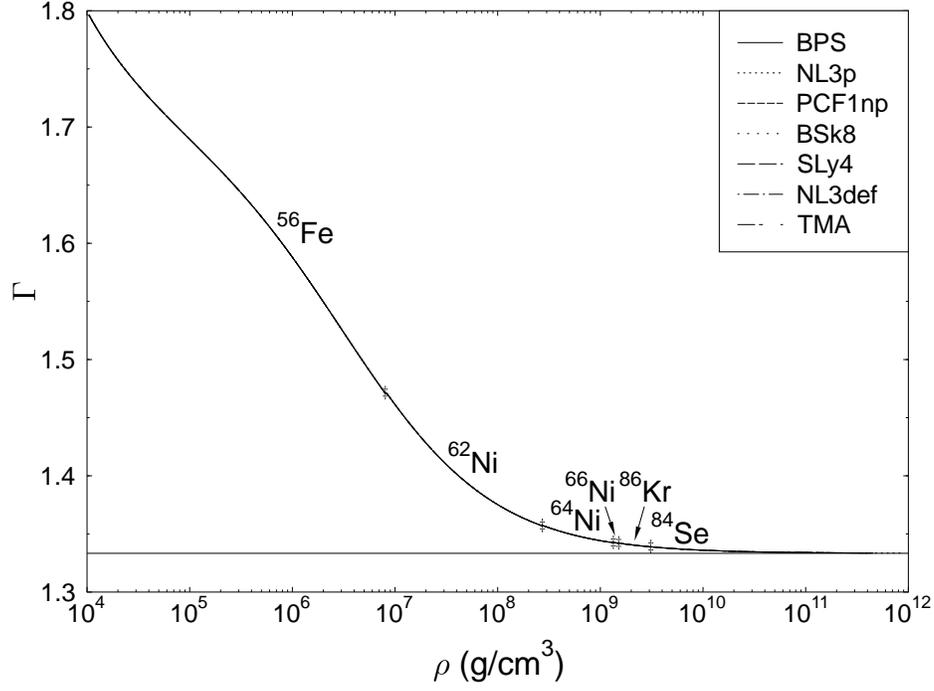}
    \caption{The adiabatic index as a
    function of the mass density, calculated by using the BPS
    model and the binding energies of various nuclear models. The
    value of $\Gamma$ of the horizontal line is equal to $\frac43$.}
    \label{adiabaticindex}
\end{figure}

\begin{figure}[ht]
    \includegraphics[scale=0.7]{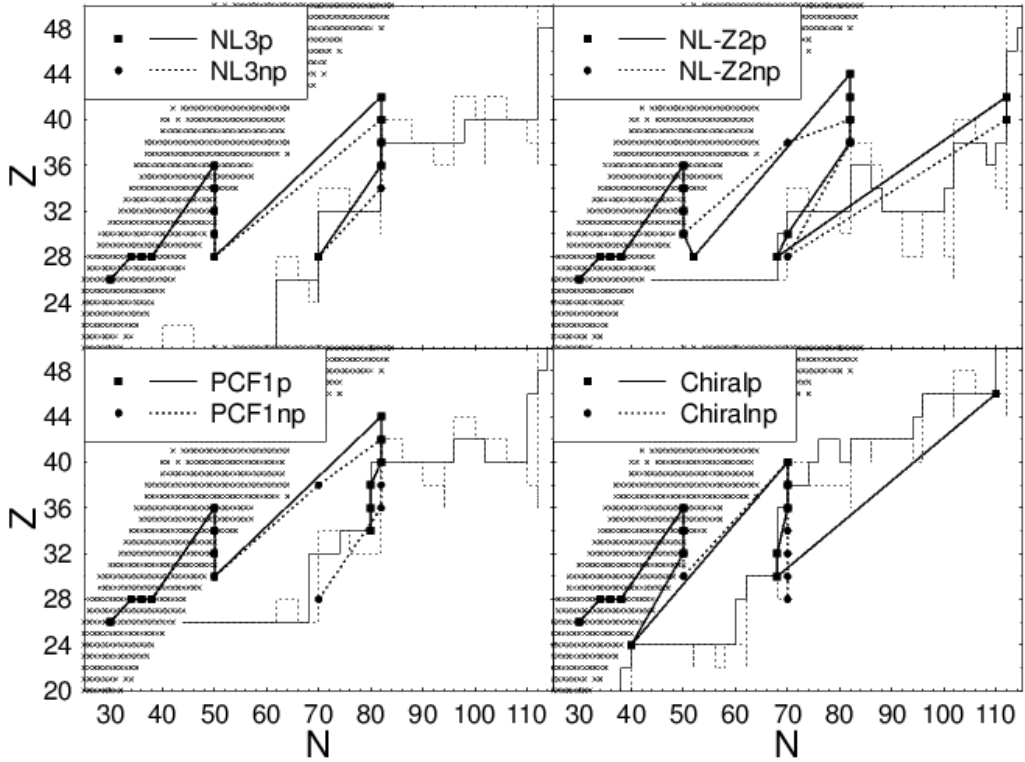}
    \caption{Nuclear charts of various relativistic nuclear models and their
    neutron driplines. The upper plots correspond to nuclear
    models with pairing while the lower ones correspond to
    nuclear models without pairing. The crosses mark the nuclei
    which are taken from the atomic mass table~\cite{AudiWapstra}. The
    thick lines and the points show the sequence of nuclei in the
    outer crust of non-accreting cold neutron stars by using
    various nuclear data.}
    \label{rmfpairing}
\end{figure}

\begin{figure}[ht]
    \includegraphics[scale=0.5]{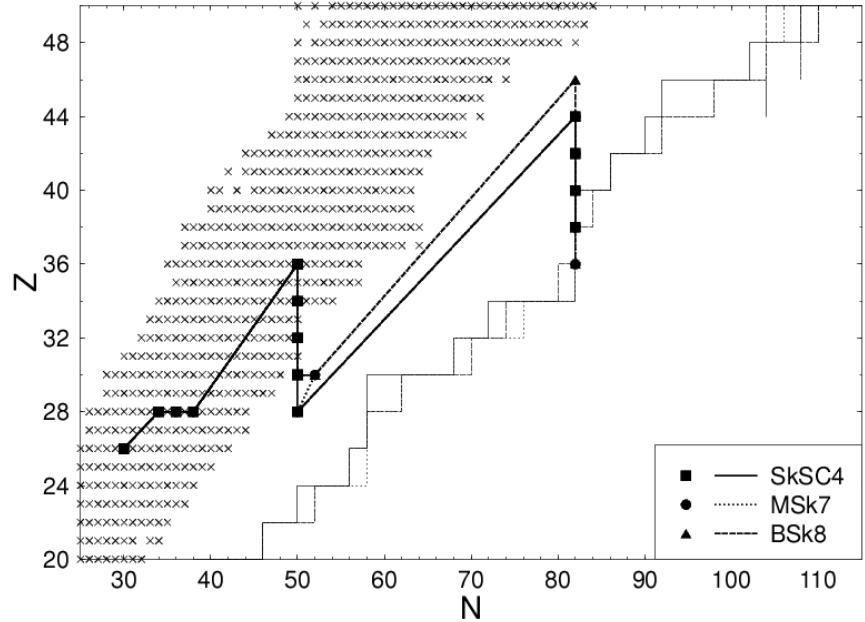}
    \caption{Nuclear chart of various non-relativistic deformed
    Skyrme Hartree-Fock-Bogolyubov calculations as taken from
    BRUSLIB~\cite{BRUSLIB_BCS,BRUSLIB_HFB} and their neutron driplines. The
    crosses mark the nuclei which are taken from the atomic mass
    table~\cite{AudiWapstra}. The thick lines and points show the sequence
    of nuclei in the outer crust of non-accreting cold neutron stars
    by using various nuclear models.}
    \label{bruslib}
\end{figure}

\begin{figure}[ht]
    \includegraphics[scale=0.5]{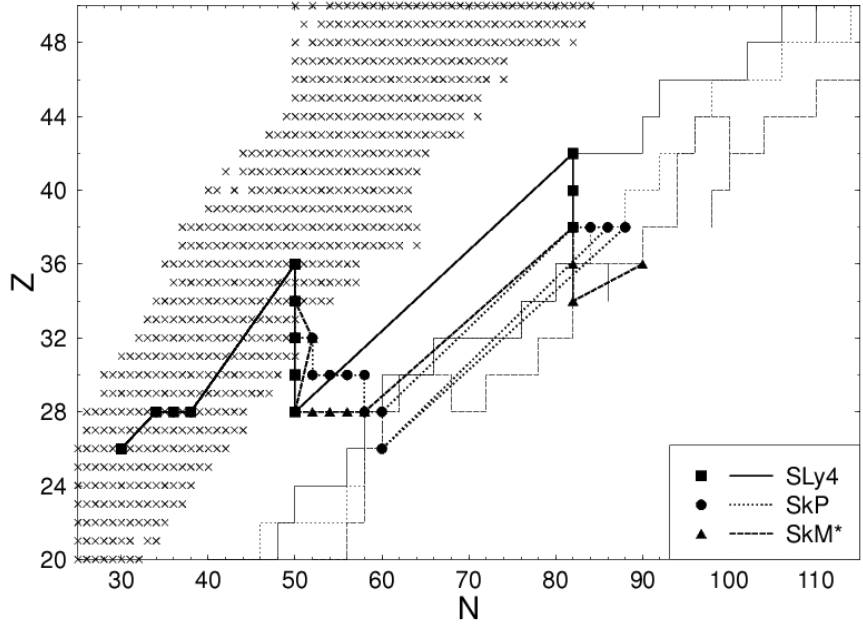}
    \caption{The same as Fig.~\ref{bruslib} but now for non-relativistic 
    deformed Skyrme Hartree-Fock-Bogolyubov models taken from
    Dobaczewski et al.~\cite{Dobaczewski}.}
    \label{dobaczewski}
\end{figure}

\begin{figure}[ht]
    \includegraphics[scale=0.55]{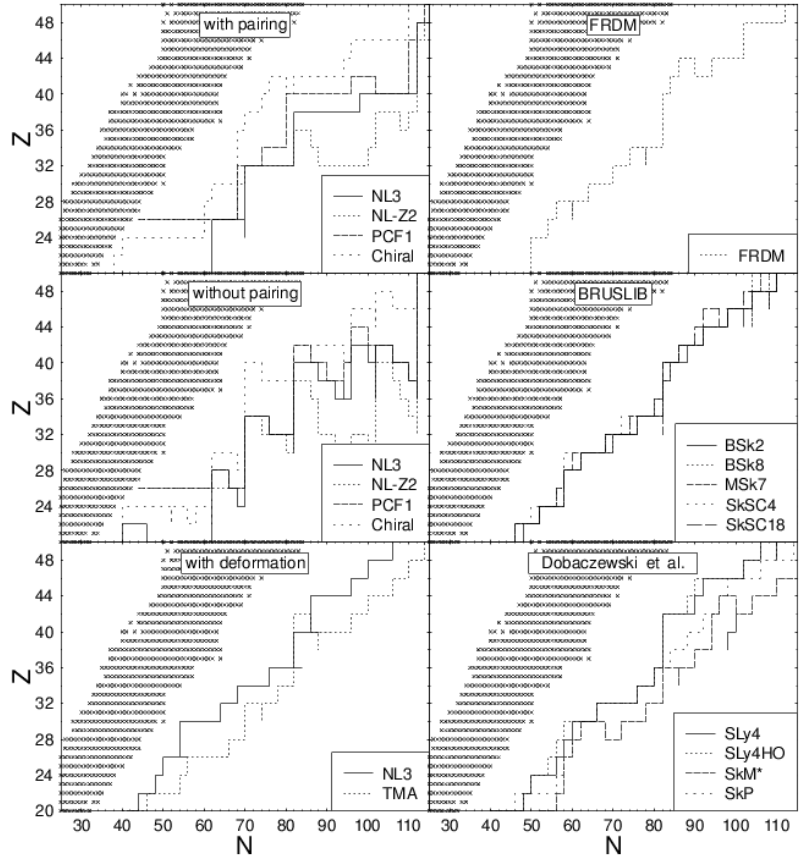}
    \caption{Nuclear charts showing the neutron driplines of
    various theoretical nuclear models (see
    Table~\ref{nuclearmodels}). The crosses mark the nuclei
    which are taken from the atomic mass table~\cite{AudiWapstra}.}
    \label{drip}
\end{figure}

\begin{figure}[ht]
    \includegraphics[scale=0.55]{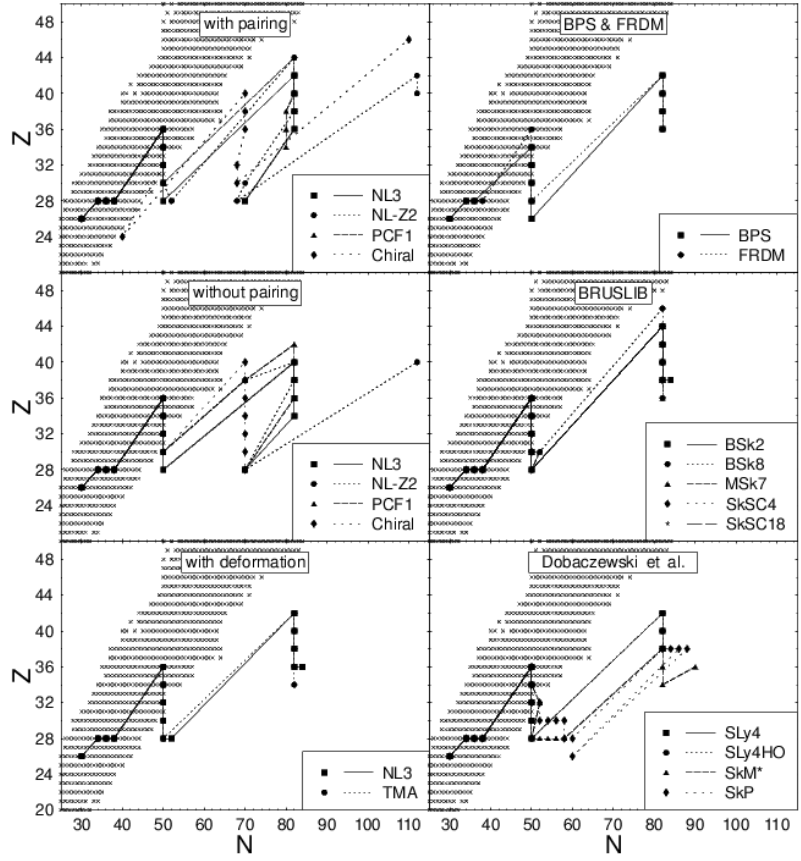}
    \caption{Same as Fig.~\ref{drip} but plotting here the sequence of
    nuclei in the outer crust of non-accreting cold neutron stars by
    using various nuclear models.}
    \label{zvonn}
\end{figure}

\begin{figure}[ht]
    \includegraphics[scale=0.48]{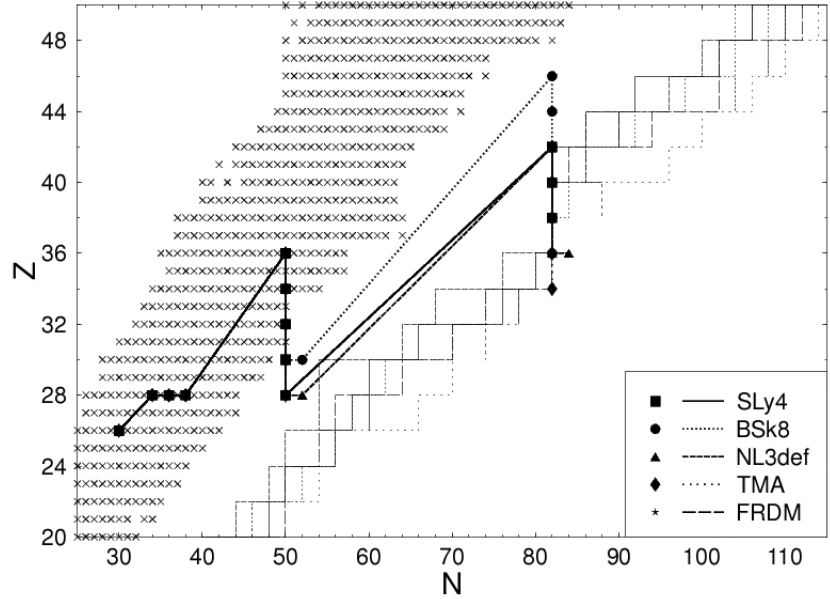}
    \caption{Nuclear chart of the selected most modern models from
    Table~\ref{nuclearmodels} and their neutron driplines. The crosses
    mark the nuclei 
    which are taken from the atomic mass table~\cite{AudiWapstra}. The
    thick lines and points show the sequence of nuclei in the
    outer crust of non-accreting cold neutron stars.}
    \label{topmodels}
\end{figure}

\begin{figure}[ht]
    \includegraphics[scale=0.48]{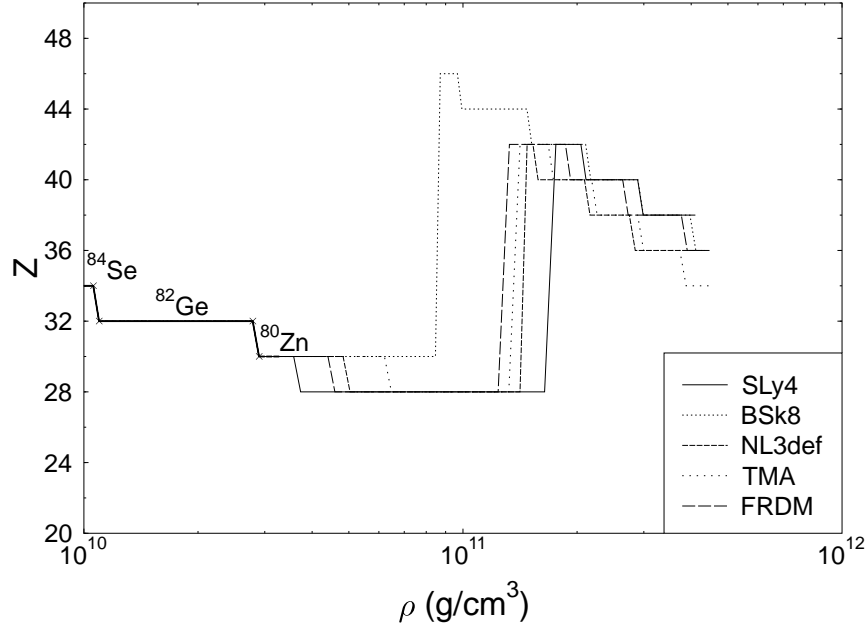}
    \caption{The proton number $Z$ as a function of the mass
    density $\rho$ for the selected most modern models as depicted also
    in Fig.~\ref{topmodels}.}
    \label{topmodelsZvonrho}
\end{figure}



\begin{table}[ht]
\begin{center}
\begin{tabular}{|c c c c c c c c|}
\hline
$\mu$ [MeV] & $\mu_e$ [MeV] & $\rho_\mathrm{max}$ [g/cm$^3$] & $P$
[dyne/cm$^2$] & $n_b$ [cm$^{-3}$] & Element & $Z$ & $N$ \\
\hline
\hline
930.60 &  0.95 & $8.09 \times 10^{6}$  & $5.29 \times 10^{23}$ & $4.88 \times 10^{30}$ &  ${}^{56}$Fe & 26 & 30 \\
931.31 &  2.60 & $2.69 \times 10^{8}$  & $6.91 \times 10^{25}$ & $1.62 \times 10^{32}$ &  ${}^{62}$Ni & 28 & 34 \\
932.00 &  4.24 & $1.24 \times 10^{9}$  & $5.20 \times 10^{26}$ & $7.48 \times 10^{32}$ &  ${}^{64}$Ni & 28 & 36 \\
933.33 &  7.69 & $8.15 \times 10^{9}$  & $5.78 \times 10^{27}$ & $4.90 \times 10^{33}$ &  ${}^{84}$Se & 34 & 50 \\
934.42 & 10.61 & $2.23 \times 10^{10}$ & $2.12 \times 10^{28}$ & $1.34 \times 10^{34}$ &  ${}^{82}$Ge & 32 & 50 \\
935.48 & 13.58 & $4.88 \times 10^{10}$ & $5.70 \times 10^{28}$ & $2.93 \times 10^{34}$ &  ${}^{80}$Zn & 30 & 50 \\
937.68 & 19.97 & $1.63 \times 10^{11}$ & $2.68 \times 10^{29}$ & $9.74 \times 10^{34}$ &  ${}^{78}$Ni & 28 & 50 \\
937.78 & 20.25 & $1.78 \times 10^{11}$ & $2.84 \times 10^{29}$ & $1.07 \times 10^{35}$ &  ${}^{76}$Fe & 26 & 50 \\
937.83 & 20.50 & $1.86 \times 10^{11}$ & $2.93 \times 10^{29}$ & $1.12 \times 10^{35}$ & ${}^{124}$Mo & 42 & 82 \\
938.57 & 22.86 & $2.67 \times 10^{11}$ & $4.55 \times 10^{29}$ & $1.60 \times 10^{35}$ & ${}^{122}$Zr & 40 & 82 \\
939.29 & 25.25 & $3.73 \times 10^{11}$ & $6.79 \times 10^{29}$ & $2.23 \times 10^{35}$ & ${}^{120}$Sr & 38 & 82 \\
939.57 & 26.19 & $4.32 \times 10^{11}$ & $7.87 \times 10^{29}$ & $2.59 \times 10^{35}$ & ${}^{118}$Kr & 36 & 82 \\
\hline
\end{tabular}
\end{center}
\caption{Sequence of nuclei in the outer crust of 
non-accreting cold neutron stars calculated by using the nuclear
data of Myers and Swiatecki~\cite{MyersSwiatecki} to 
reproduce the results of BPS~\cite{BPS}. The last line corresponds
to the neutron drip point.}
\label{BPStable}
\end{table}

\begin{table}[ht]
\begin{center}
\begin{tabular}{|c c c c c|}
\hline
$\mu_e$ [MeV] & $\rho_\mathrm{max}$ [g/cm$^3$] & Element & $Z$ & $N$ \\
\hline
\hline
0.95  & $7.96 \times 10^{6}$  & ${}^{56}$Fe  & 26 & 56 \\
2.61  & $2.70 \times 10^{8}$  & ${}^{62}$Ni  & 28 & 34 \\
4.17  & $1.18 \times 10^{9}$  & ${}^{64}$Ni  & 28 & 36 \\
6.94  & $5.88 \times 10^{9}$  & ${}^{68}$Ni  & 28 & 40 \\
9.12  & $1.36 \times 10^{10}$ & ${}^{84}$Se  & 34 & 50 \\
9.16  & $1.40 \times 10^{10}$ & ${}^{70}$Ni  & 28 & 42 \\
10.06 & $1.93 \times 10^{10}$ & ${}^{72}$Ni  & 28 & 44 \\
10.86 & $2.43 \times 10^{10}$ & ${}^{78}$Zn  & 30 & 48 \\
13.24 & $4.52 \times 10^{10}$ & ${}^{80}$Zn  & 30 & 50 \\
13.58 & $4.97 \times 10^{10}$ & ${}^{76}$Ni  & 28 & 48 \\
16.51 & $9.17 \times 10^{10}$ & ${}^{78}$Ni  & 28 & 50 \\
18.04 & $1.23 \times 10^{11}$ & ${}^{80}$Ni  & 28 & 52 \\
19.38 & $1.56 \times 10^{11}$ & ${}^{82}$Ni  & 28 & 54 \\
21.12 & $2.07 \times 10^{11}$ & ${}^{84}$Ni  & 28 & 56 \\
22.89 & $2.70 \times 10^{11}$ & ${}^{86}$Ni  & 28 & 58 \\
24.36 & $3.33 \times 10^{11}$ & ${}^{88}$Ni  & 28 & 60 \\
25.10 & $3.77 \times 10^{11}$ & ${}^{130}$Zr & 40 & 90 \\
25.97 & $4.25 \times 10^{11}$ & ${}^{132}$Zr & 40 & 92 \\
26.26 & $4.45 \times 10^{11}$ & ${}^{134}$Zr & 40 & 94 \\
\hline
\end{tabular}
\end{center}
\caption{Sequence of nuclei in the outer crust of 
non-accreting cold neutron stars as calculated by Haensel, Zdunik,
and Dobaczewski~\cite{HZD}. Nuclear masses are taken from a spherical
calculation using the parameter set SkP of Dobaczewski, Flocard, and
Treiner~\cite{HFBHZD}.}
\label{HZDtable1}
\end{table}

\begin{table}[ht]
\begin{center}
\begin{tabular}{|c c c c c|}
\hline
$\mu_e$ [MeV] & $\rho_\mathrm{max}$ [g/cm$^3$] & Element & $Z$ & $N$ \\
\hline
\hline
0.95  & $7.96 \times 10^{6}$  & ${}^{56}$Fe  & 26 & 30 \\
2.61  & $2.70 \times 10^{8}$  & ${}^{62}$Ni  & 28 & 34 \\
4.28  & $1.29 \times 10^{9}$  & ${}^{64}$Ni  & 28 & 36 \\
4.57  & $1.61 \times 10^{9}$  & ${}^{66}$Ni  & 28 & 38 \\
5.32  & $2.63 \times 10^{9}$  & ${}^{68}$Ni  & 28 & 40 \\
6.21  & $4.34 \times 10^{9}$  & ${}^{80}$Ge  & 32 & 48 \\
9.69  & $1.70 \times 10^{10}$ & ${}^{82}$Ge  & 32 & 50 \\
12.26 & $3.59 \times 10^{10}$ & ${}^{80}$Zn  & 30 & 50 \\
18.22 & $1.23 \times 10^{11}$ & ${}^{78}$Ni  & 28 & 50 \\
18.73 & $1.41 \times 10^{11}$ & ${}^{76}$Fe  & 26 & 50 \\
20.15 & $1.83 \times 10^{11}$ & ${}^{122}$Zr & 40 & 82 \\
22.19 & $2.53 \times 10^{11}$ & ${}^{120}$Sr & 38 & 82 \\
24.24 & $3.42 \times 10^{11}$ & ${}^{118}$Kr & 36 & 82 \\
26.28 & $4.55 \times 10^{11}$ & ${}^{116}$Se & 34 & 82 \\
26.82 & $5.05 \times 10^{11}$ & ${}^{114}$Ge & 32 & 82 \\
\hline
\end{tabular}
\end{center}
\caption{Sequence of nuclei in the outer crust of 
non-accreting cold neutron stars calculated by Haensel, Zdunik,
and Dobaczewski~\cite{HZD}. Nuclear masses are taken from the
droplet model of Myers~\cite{Myers}.}
\label{HZDtable2}
\end{table}

\begin{table}[ht]
\begin{center}
\begin{tabular}{|c c c c c|}
\hline
$\mu_e$ [MeV] & $\rho_\mathrm{max}$ [g/cm$^3$] & Element & $Z$ & $N$ \\
\hline
\hline
 0.95 & $7.96 \times 10^{6}$  &  ${}^{56}$Fe & 26 & 30 \\
 2.61 & $2.71 \times 10^{8}$  &  ${}^{62}$Ni & 28 & 34 \\
 4.31 & $1.30 \times 10^{9}$  &  ${}^{64}$Ni & 28 & 36 \\
 4.45 & $1.48 \times 10^{9}$  &  ${}^{66}$Ni & 28 & 38 \\
 5.66 & $3.12 \times 10^{9}$  &  ${}^{86}$Kr & 36 & 50 \\
 8.49 & $1.10 \times 10^{10}$ &  ${}^{84}$Se & 34 & 50 \\
11.44 & $2.80 \times 10^{10}$ &  ${}^{82}$Ge & 32 & 50 \\
14.08 & $5.44 \times 10^{10}$ &  ${}^{80}$Zn & 30 & 50 \\
16.78 & $9.64 \times 10^{10}$ &  ${}^{78}$Ni & 28 & 50 \\ 
\hline
18.34 & $1.29 \times 10^{11}$ & ${}^{126}$Ru & 44 & 82 \\
20.56 & $1.88 \times 10^{11}$ & ${}^{124}$Mo & 42 & 82 \\
22.86 & $2.67 \times 10^{11}$ & ${}^{122}$Zr & 40 & 82 \\
25.38 & $3.79 \times 10^{11}$ & ${}^{120}$Sr & 38 & 82 \\
26.19 & $4.33 \times 10^{11}$ & ${}^{118}$Kr & 36 & 82 \\
\hline
\end{tabular}
\end{center}
\caption{Sequence of nuclei in the outer crust of 
non-accreting cold neutron stars calculated by Haensel and
Pichon~\cite{HP}. Upper part: using experimental nuclear masses. Lower part:
using binding energies from the mass formula of M\"oller and
Nix~\cite{Moeller}. The last line corresponds to the neutron drip point.}
\label{HaenselPichontable}
\end{table}

\begin{table}[ht]
\begin{center}
\begin{tabular}{|c c c c c|}
\hline
$\mu_e$ [MeV] & $\rho_\mathrm{max}$ [g/cm$^3$] & Element & $Z$ & $N$ \\
\hline
\hline
17.44 & $1.08 \times 10^{11}$ &  ${}^{78}$Ni & 28 & 50 \\
\hline
19.13 & $1.47 \times 10^{11}$ & ${}^{126}$Ru & 44 & 82 \\
21.66 & $2.20 \times 10^{11}$ & ${}^{124}$Mo & 42 & 82 \\
24.13 & $3.15 \times 10^{11}$ & ${}^{122}$Zr & 40 & 82 \\
26.05 & $4.10 \times 10^{11}$ & ${}^{120}$Sr & 38 & 82 \\
\hline
\end{tabular}
\end{center}
\caption{Sequence of nuclei in the outer crust of 
non-accreting cold neutron stars calculated by Haensel and
Pichon~\cite{HP}. Upper part: last experimental nuclear mass. Lower part:
using the mass formula of Aboussir et al.~\cite{Aboussir}.
The last line corresponds to the neutron drip point.}
\label{HaenselPichontable2}
\end{table}

\begin{table}[ht]
\begin{center}
\begin{tabular}{|c c c c c c c c|}
\hline
$\mu$ [MeV] & $\mu_e$ [MeV] & $\rho_\mathrm{max}$ [g/cm$^3$] & $P$
[dyne/cm$^2$] & $n_b$ [cm$^{-3}$] & Element & $Z$ & $N$ \\
\hline
\hline
930.60 &  0.95 & $8.02 \times 10^{6}$  & $5.22 \times 10^{23}$ & $4.83 \times 10^{30}$ &  ${}^{56}$Fe & 26 & 30 \\
931.32 &  2.61 & $2.71 \times 10^{8}$  & $6.98 \times 10^{25}$ & $1.63 \times 10^{32}$ &  ${}^{62}$Ni & 28 & 34 \\
932.04 &  4.34 & $1.33 \times 10^{9}$  & $5.72 \times 10^{26}$ & $8.03 \times 10^{32}$ &  ${}^{64}$Ni & 28 & 36 \\
932.09 &  4.46 & $1.50 \times 10^{9}$  & $6.44 \times 10^{26}$ & $9.04 \times 10^{32}$ &  ${}^{66}$Ni & 28 & 38 \\
932.56 &  5.64 & $3.09 \times 10^{9}$  & $1.65 \times 10^{27}$ & $1.86 \times 10^{33}$ &  ${}^{86}$Kr & 36 & 50 \\
933.62 &  8.38 & $1.06 \times 10^{10}$ & $8.19 \times 10^{27}$ & $6.37 \times 10^{33}$ &  ${}^{84}$Se & 34 & 50 \\
934.75 & 11.43 & $2.79 \times 10^{10}$ & $2.85 \times 10^{28}$ & $1.68 \times 10^{34}$ &  ${}^{82}$Ge & 32 & 50 \\
935.89 & 14.61 & $6.07 \times 10^{10}$ & $7.63 \times 10^{28}$ & $3.65 \times 10^{34}$ &  ${}^{80}$Zn & 30 & 50 \\
\hline
936.44 & 16.17 & $8.46 \times 10^{10}$ & $1.15 \times 10^{29}$ & $5.08 \times 10^{34}$ &  ${}^{82}$Zn & 30 & 52 \\
936.63 & 16.81 & $9.67 \times 10^{10}$ & $1.32 \times 10^{29}$ & $5.80 \times 10^{34}$ & ${}^{128}$Pd & 46 & 82 \\
937.41 & 19.16 & $1.47 \times 10^{11}$ & $2.23 \times 10^{29}$ & $8.84 \times 10^{34}$ & ${}^{126}$Ru & 44 & 82 \\
938.12 & 21.35 & $2.11 \times 10^{11}$ & $3.45 \times 10^{29}$ & $1.26 \times 10^{35}$ & ${}^{124}$Mo & 42 & 82 \\
938.78 & 23.47 & $2.89 \times 10^{11}$ & $5.05 \times 10^{29}$ & $1.73 \times 10^{35}$ & ${}^{122}$Zr & 40 & 82 \\
939.47 & 25.77 & $3.97 \times 10^{11}$ & $7.36 \times 10^{29}$ & $2.38 \times 10^{35}$ & ${}^{120}$Sr & 38 & 82 \\
939.57 & 26.09 & $4.27 \times 10^{11}$ & $7.74 \times 10^{29}$ & $2.56 \times 10^{35}$ & ${}^{118}$Kr & 36 & 82 \\
\hline
\end{tabular}
\end{center}
\caption{Sequence of nuclei in the outer crust of 
non-accreting cold neutron stars calculated by using the
experimental nuclear data from the atomic mass
table~\cite{AudiWapstra} (upper part), and the theoretical mass
table of the Skyrme model BSk8 as listed by BRUSLIB (lower part).
Note that the experimental data from
the atomic mass table~\cite{AudiWapstra} is always taken first if
available. The last line corresponds to the neutron drip point.}
\label{BSk8table}
\end{table}

\begin{table}[ht]
\begin{center}
\begin{tabular}{|c c c c c c c c|}
\hline
$\mu$ [MeV] & $\mu_e$ [MeV] & $\rho_\mathrm{max}$ [g/cm$^3$] & $P$
[dyne/cm$^2$] & $n_b$ [cm$^{-3}$] & Element & $Z$ & $N$ \\
\hline
\hline
930.60 &  0.95 & $8.02 \times 10^{6}$  & $5.22 \times 10^{23}$ & $4.83 \times 10^{30}$ &  ${}^{56}$Fe & 26 & 30 \\
931.32 &  2.61 & $2.71 \times 10^{8}$  & $6.98 \times 10^{25}$ & $1.63 \times 10^{32}$ &  ${}^{62}$Ni & 28 & 34 \\
932.04 &  4.34 & $1.33 \times 10^{9}$  & $5.72 \times 10^{26}$ & $8.03 \times 10^{32}$ &  ${}^{64}$Ni & 28 & 36 \\
932.09 &  4.46 & $1.50 \times 10^{9}$  & $6.44 \times 10^{26}$ & $9.04 \times 10^{32}$ &  ${}^{66}$Ni & 28 & 38 \\
932.56 &  5.64 & $3.09 \times 10^{9}$  & $1.65 \times 10^{27}$ & $1.86 \times 10^{33}$ &  ${}^{86}$Kr & 36 & 50 \\
933.62 &  8.38 & $1.06 \times 10^{10}$ & $8.19 \times 10^{27}$ & $6.37 \times 10^{33}$ &  ${}^{84}$Se & 34 & 50 \\
934.75 & 11.43 & $2.79 \times 10^{10}$ & $2.85 \times 10^{28}$ & $1.68 \times 10^{34}$ &  ${}^{82}$Ge & 32 & 50 \\
935.93 & 14.71 & $6.21 \times 10^{10}$ & $7.86 \times 10^{28}$ & $3.73 \times 10^{34}$ &  ${}^{80}$Zn & 30 & 50 \\
\hline
937.28 & 18.64 & $1.32 \times 10^{11}$ & $2.03 \times 10^{29}$ & $7.92 \times 10^{34}$ &  ${}^{78}$Ni & 28 & 50 \\
937.63 & 19.80 & $1.68 \times 10^{11}$ & $2.55 \times 10^{29}$ & $1.01 \times 10^{35}$ & ${}^{124}$Mo & 42 & 82 \\
938.13 & 21.38 & $2.18 \times 10^{11}$ & $3.48 \times 10^{29}$ & $1.31 \times 10^{35}$ & ${}^{122}$Zr & 40 & 82 \\
938.67 & 23.19 & $2.89 \times 10^{11}$ & $4.82 \times 10^{29}$ & $1.73 \times 10^{35}$ & ${}^{120}$Sr & 38 & 82 \\
939.18 & 24.94 & $3.73 \times 10^{11}$ & $6.47 \times 10^{29}$ & $2.23 \times 10^{35}$ & ${}^{118}$Kr & 36 & 82 \\
939.57 & 26.29 & $4.55 \times 10^{11}$ & $8.00 \times 10^{29}$ & $2.72 \times 10^{35}$ & ${}^{116}$Se & 34 & 82 \\
\hline
\end{tabular}
\end{center}
\caption{The same as Table~\ref{BSk8table} but for the
relativistic nuclear model TMA.}
\label{TMAtable}
\end{table}

\begin{table}[ht]
\begin{center}
\begin{tabular}{|c c c c l l|}
\hline
Year & Element & Z & N & Model & Refs. \\
\hline
\hline
1966 & ${}^{118}$Kr & 36 & 82 & Droplet model of Myers and Swiatecki & \cite{MyersSwiatecki} \\
1977 & ${}^{114}$Ge & 32 & 82 & Droplet model of Myers & \cite{Myers} \\
1984 & ${}^{134}$Zr & 40 & 94 & Skyrme model SkP (spherical) & \cite{HFBHZD} \\
1988 & ${}^{118}$Kr & 36 & 82 & Droplet model of M\"oller and Nix & \cite{Moeller} \\
1992 & ${}^{120}$Sr & 38 & 82 & Droplet model of Aboussir et al. & \cite{Aboussir} \\
\hline
1995      & ${}^{118}$Kr & 36 & 82 & Finite range droplet model FRDM & \cite{Moller97} \\
1997/1999 & ${}^{120}$Kr & 36 & 84 & NL3 (with deformations) & \cite{Lala97,Lala99} \\
1998/2004 & ${}^{120}$Sr & 38 & 82 & SLy4 (with deformations) & \cite{HaenselSLy4,Dobaczewski} \\
2004      & ${}^{118}$Kr & 36 & 82 & BSk8 (with deformations) & \cite{BRUSLIB_HFB} \\
2005      & ${}^{116}$Se & 34 & 82 & TMA (with deformations) & \cite{Geng05} \\
\hline
\end{tabular}
\end{center}
\caption{The endpoint of the sequence of nuclei obtained by 
using different theoretical nuclear models. Upper part: results
obtained in previous work by BPS, HZD, and HP with older nuclear
models. Lower part: results obtained in this work by using modern 
models and mass tables (see Table~\ref{nuclearmodels}).} 
\label{last}
\end{table}


\end{widetext}


\begin{thebibliography}{99}

\bibitem{FMT} R. P. Feynman, N. Metropolis, and E. Teller,
Phys. Rev. \textbf{75}, 1561 (1949).

\bibitem{ocean} E. F. Brown, and L. Bildsten, Ap. J.
\textbf{496}, 915 (1998).

\bibitem{Ravenhall} C. J. Pethick, and D. G. Ravenhall,
Annu. Rev. Nucl. Part. Sci. \textbf{45}, 429 (1995).

\bibitem{BPS} G. Baym, C. Pethick, and P. Sutherland,
Ap. J. \textbf{170}, 299 (1971).

\bibitem{BBP} G. Baym, H. A. Bethe, and C. J. Pethick,
Nucl. Phys. \textbf{A175}, 225 (1971).

\bibitem{NV} J. W. Negele, and D. Vautherin,
Nucl. Phys. \textbf{A207}, 298 (1973).

\bibitem{Shen}
H. Shen, Phys. Rev. C \textbf{65}, 035802 (2002).

\bibitem{Akmal98}
B. Friedman, and V. R. Pandharipande,
Nucl. Phys. \textbf{A361}, 502 (1981);
A. Akmal, V. R. Pandharipande, and D. G. Ravenhall,
Phys. Rev. C \textbf{58}, 1804 (1998);
M. Baldo, G. F. Burgio, and H. J. Schulze,
Phys. Rev. C \textbf{61}, 055801 (2000);
I. Vidana, A. Polls, A. Ramos, L. Engvik, and M. Hjorth-Jensen,
Phys. Rev. C \textbf{62}, 035801 (2000).

\bibitem{Walecka}
B. D. Serot, and J. D. Walecka, 
Advances in Nuclear Physics \textbf{16}, 
ed. J. W. Negele, and E. Vogt (Plenum, NY, 1986); 
B. D. Serot, Rep. Phys. \textbf{55}, 1855 (1992).

\bibitem{RMF}
N. K. Glendenning, Ap. J. \textbf{293}, 470 (1985);
P. J. Ellis, R. Knorren, and M. Prakash,
Phys. Lett. B \textbf{349}, 11 (1995);
J. Schaffner, and I. N. Mishustin, 
Phys. Rev. C \textbf{53}, 1416 (1996);
R. Knorren, M. Prakash, and P. J. Ellis,
Phys. Rev. C \textbf{52}, 3470 (1995);
H. Huber, F. Weber, M. K. Weigel, and C. Schaab,
Int. J. Mod. Phys. E \textbf{7}, 301 (1998);
S. Pal, M. Hanauske, I. Zakout, H. St\"ocker, and W. Greiner,
Phys. Rev. C \textbf{60}, 015802 (1999);
J. Schaffner-Bielich, M. Hanauske, H. St\"ocker, and W. Greiner,
Phys. Rev. Lett. \textbf{89}, 171101 (2002).

\bibitem{Hanauske00}
M. Hanauske, D. Zschiesche, S. Pal, S. Schramm, H. St\"ocker,
W. Greiner, Ap. J. \textbf{537}, 958 (2000).

\bibitem{SZ02}
S. Schramm, and D. Zschiesche, J. Phys. G \textbf{29}, 531 (2003).

\bibitem{Weber}
F. Weber, Prog. Part. Nucl. Phys. \textbf{54}, 193 (2005).

\bibitem{CSCReviews} D. H. Rischke, Prog. Part. Nucl. Phys.
\textbf{52}, 197 (2004); K. Rajagopal, and F. Wilzcek,
hep-ph/0011333; M. Alford, Ann. Rev. Nucl. Part. Sci.
\textbf{51}, (2001).

\bibitem{QCDphasediagrams} S. B. R\"uster, I. A. Shovkovy, and
D. H. Rischke, Nucl. Phys. \textbf{A743}, 127 (2004);
K. Fukushima, C. Kouvaris, and K. Rajagopal, Phys. Rev.
D \textbf{71}, 034002 (2005);
I. A. Shovkovy, S. B. R\"uster, and D. H. Rischke, J. Phys. G:
Nucl. Part. Phys. \textbf{31}, S849 (2005);
S. B. R\"uster, V. Werth, M. Buballa, I. A. Shovkovy, and
D. H. Rischke, Phys. Rev. D \textbf{72}, 034004 (2005);
D. Blaschke, S. Fredriksson, H. Grigorian, A. M. \"Ozta\c s, and F.
Sandin, hep-ph/0503194;
S. B. R\"uster, V. Werth, M. Buballa, I. A. Shovkovy, and
D. H. Rischke, hep-ph/0509073.

\bibitem{MyersSwiatecki} W. D. Myers, and W. J. Swiatecki, UCRL
Report 11980 (unpublished); W. D. Myers, and W. J. Swiatecki,
Nucl. Phys. \textbf{81}, 1 (1966).

\bibitem{HZD} P. Haensel, J. L. Zdunik, and J. Dobaczewski,
Astron. Astrophys. \textbf{222}, 353 (1989).

\bibitem{HFBHZD} J. Dobaczewski, H. Flocard, and J. Treiner,
Nucl. Phys. \textbf{A422}, 103 (1984).

\bibitem{Myers} W. D. Myers, \textit{Droplet Model of Atomic
Nucleus}, Plenum, New York (1977).

\bibitem{HP} P. Haensel, and B. Pichon,
Astron. Astrophys. \textbf{283}, 313 (1994).

\bibitem{AW1992} G. Audi, and A. H. Wapstra, Nucl. Phys.
\textbf{A565}, 1 (1993).

\bibitem{Moeller} P. M\"oller, and J. R. Nix, Atom. Data Nucl.
Data Tables \textbf{39}, 213 (1988).

\bibitem{Aboussir} Y. Aboussir, J. M. Pearson, A. K. Dutta, and
F. Tondeur, Nucl. Phys. \textbf{A549}, 155 (1992).

\bibitem{ReviewHaensel} P. Haensel, \textit{Neutron Star Crusts},
\textit{Physics of Neutron Star Interiors}, Lecture Notes in Physics
\textbf{578}, Springer Verlag Heidelberg, 127 (2001).

\bibitem{AudiWapstra} G. Audi, A. H. Wapstra, and C. Thibault,
Nucl. Phys. \textbf{A729}, 337 (2003).

\bibitem{BRUSLIB_BCS} F. Tondeur, S. Goriely, J. M. Pearson,
and M. Onsi, Phys. Rev. C \textbf{62}, 024308 (2000); 
S. Goriely, J. M. Pearson, and F. Tondeur, At. Data and
Nucl. Data Tables \textbf{77}, 311 (2001);
Nuclear data downloaded from BRUSLIB:
http://www-astro.ulb.ac.be/Html/masses.html.

\bibitem{BRUSLIB_HFB}
M. Samyn, S. Goriely, P.-H. Heenen, J. M. Pearson, and F. Tondeur,
Nucl. Phys. \textbf{A700}, 142 (2001); 
S. Goriely, M. Samyn, P.-H. Heenen, J. M. Pearson, and F. Tondeur,
Phys. Rev. C \textbf{66}, 024326 (2002); 
M. Samyn, S. Goriely, and J. M. Pearson, 
Nucl. Phys. \textbf{A725}, 69 (2003); 
S. Goriely, M. Samyn, M. Bender, and J. M. Pearson,
Phys. Rev. C \textbf{68}, 054325 (2003); 
M. Samyn, S. Goriely, M. Bender, and J. M. Pearson,
Phys. Rev. C \textbf{70}, 044309 (2004);
S. Goriely, M. Samyn, J. M. Pearson, and M. Onsi,
Nucl. Phys. \textbf{A750}, 425 (2005).
Nuclear data downloaded from BRUSLIB:
http://www-astro.ulb.ac.be/Html/masses.html.

\bibitem{Dobaczewski} M. V. Stoitsov, J. Dobaczewski, W. Nazarewicz, 
S. Pittel, and D. J. Dean, Phys. Rev. C \textbf{68}, 054312
(2003); J. Dobaczewski, M. V. Stoitsov, W. Nazarewicz, 
AIP Conference Proceedings Volume \textbf{726}, ed. R. Bijker, R. F. Casten, 
A. Frank (American Institute of Physics, New York, 2004) p. 51,
nucl-th/0404077;
Nuclear data downloaded from:
http://www.fuw.edu.pl/$\sim$dobaczew/thodri/thodri.html.

\bibitem{Geng05}
L. S. Geng, H. Toki, and J. Meng, Prog. Theor. Phys.
\textbf{113}, 785 (2005);
Lisheng Geng, private communications.

\bibitem{FAIR} 
http://www.gsi.de/fair.

\bibitem{TRIUMF} 
http://www.triumf.info.

\bibitem{RIA} 
http://www.orau.org/ria.

\bibitem{CHM} R. A. Coldwell-Horsfall, and A. A. Maradudin,
J. Math. Phys. \textbf{1}, 395 (1960).

\bibitem{Moller97}
P. M\"oller, J. R. Nix, W. D. Myers, and W. J. Swiatecki, 
Atom. Data Nucl. Data Tabl. \textbf{59}, 185 (1995);
P. M\"oller, J. R. Nix, and K. L. Kratz,
Atom. Data Nucl. Data Tabl. \textbf{66}, 131 (1997).

\bibitem{Bartel82}
J. Bartel, P. Quentin, M. Brack, C. Guet, and H. B. Hakansson,
Nucl. Phys. \textbf{A386}, 79 (1982).

\bibitem{Doba84}
J. Dobaczewski, H. Flocard, and J. Treiner,
Nucl. Phys. \textbf{A422}, 103 (1984).

\bibitem{HaenselSLy4} E. Chabanat, P. Bonche, P. Haensel, J.
Meyer, and R. Schaeffer, Nucl. Phys. \textbf{A627}, 710 (1997);
Nucl. Phys. \textbf{A635}, 231 (1998), 
erratum: ibid. \textbf{A643}, 441 (1998).

\bibitem{Kri80}
H. Krivine, J. Treiner, O. Bohigas,
Nucl. Phys. \textbf{A336}, 155 (1980).

\bibitem{Wiringa88}
R. B. Wiringa, V. Fiks, and A. Fabrocini,
Phys. Rev. C \textbf{38}, 1010 (1988).

\bibitem{Papa99}
P. Papazoglou, D. Zschiesche, S. Schramm, J. Schaffner-Bielich,
H. St\"ocker, and W. Greiner, 
Phys. Rev. C \textbf{59}, 411 (1999).

\bibitem{Beckmann02}
C. Beckmann, P. Papazoglou, D. Zschiesche, S. Schramm, H.
St\"ocker, and W. Greiner, 
Phys. Rev. C \textbf{65}, 024301 (2002).

\bibitem{Schramm02}
S. Schramm,
Phys. Rev. C \textbf{66}, 064310 (2002).

\bibitem{Schramm03}
S. Schramm,
Phys. Lett. B \textbf{560}, 164 (2003).

\bibitem{Lala99}
G. A. Lalazissis, S. Raman, P. Ring,
Atom. Data Nucl. Data Tabl. \textbf{71}, 1 (1999).

\bibitem{Lala97}
G. A. Lalazissis, J. K\"onig, and P. Ring,
Phys. Rev. C \textbf{55}, 540 (1997).

\bibitem{Bender99}
M. Bender, K. Rutz, P. G. Reinhard, J. A. Maruhn, and W. Greiner,
Phys. Rev. C \textbf{60}, 034304 (1999).

\bibitem{Buervenich02}
T. B\"urvenich, D. G. Madland, J. A. Maruhn, P. G. Reinhard,
Phys. Rev. C \textbf{65}, 044308 (2002).

\bibitem{Suga94}
Y. Sugahara, and H. Toki,
Nucl. Phys. A \textbf{579}, 557 (1994).

\bibitem{Cottam2002}
J. Cottam, F. Paerels, and M. Mendez, Nature \textbf{420}, 51 (2002).

\end{thebibliography}
\end{document}